\documentclass[fleqn,usenatbib]{mnras}

\usepackage{newtxtext,newtxmath}
\usepackage[T1]{fontenc}

\DeclareRobustCommand{\VAN}[3]{#2}
\let\VANthebibliography\thebibliography
\def\thebibliography{\DeclareRobustCommand{\VAN}[3]{##3}\VANthebibliography}

\usepackage{graphicx}	% Including figure files
\usepackage{amsmath}	% Advanced maths commands
\usepackage{gensymb}    % Generic symbols

\usepackage{xcolor}
\usepackage{verbatim}

\newcommand\ms{\ensuremath{\text{m}\,\text{s}^{-1}}} 			%print metres per second
\newcommand\masyr{\ensuremath{\text{mas}\,\text{yr}^{-1}}} 		%print mas per year

\definecolor{my_color}{HTML}{CF0000}                            %this is a maroon
\newcommand\bmaroon{\textcolor{black}}                          %defunct
\newcommand\boldnew{\textcolor{black}}                          %defunct
\newcommand\boldagain{\textcolor{black}}                        %defunct

\title[Edge-On Orbit for HR 5183 b]{An Edge-On Orbit for the Eccentric Long-Period Planet HR 5183 b}

\author[Venner et al.]{Alexander Venner,$^{1}$\thanks{E-mail: AlexanderVenner@gmail.com}
Logan A. Pearce,$^{2}$\thanks{NSF Graduate Research Fellow}
Andrew Vanderburg$^{3}$
\\
$^{1}$Aberdeen, UK\\
$^{2}$Steward Observatory, University of Arizona, Tucson, AZ 85721, USA\\
$^{3}$Department of Physics and Kavli Institute for Astrophysics and Space Research, Massachusetts Institute of Technology, Cambridge, MA 02139, USA\\
}

\date{Accepted 2022 August 23. Received 2022 August 12; in original form 2021 November 5}

\pubyear{2022}

\begin{document}
\label{firstpage}
\pagerange{\pageref{firstpage}--\pageref{lastpage}}
\maketitle

\begin{abstract}

The long-period giant planet HR 5183 b has one of the most extreme orbits among exoplanets known to date\boldnew{, and represents a test for models of their dynamical evolution}. In this work we use Hipparcos-Gaia astrometry to measure the orbital inclination of this planet for the first time \boldnew{and find} $i=89.9^{+13.3\circ}_{-13.5}$, fully consistent with \boldnew{edge-on. The} long orbital period and high eccentricity of HR 5183 b are supported by our results, with $P=102^{+84}_{-34}$ years and $e=0.87 \pm 0.04$. We confirm that \boldnew{HR 5183} forms a physically bound binary with \boldnew{HIP 67291} at a projected separation of 15400 AU, and derive new constraints on the orbit of this pair. We combine \boldnew{these results} to measure the mutual inclination between the planetary and binary orbits; \boldnew{we observe significant evidence for misalignment, which remains even after accounting for bias of the prior towards high mutual inclinations. However, our results are too imprecise to evaluate a recent prediction that the mutual inclination should reflect the formation history of HR 5183 b.} Further observations, especially the release of the full Gaia astrometric data, will allow for improved constraints on the planet-binary mutual inclination. $52 \pm 16\%$ of known planets with eccentricities $e\geq 0.8$ are found in multiple star systems, a rate that we find to be greater than for the overall planet population to moderate significance ($p=0.0075$). This supports the hypothesis that dynamical interactions with wide stellar companions plays an important role in the formation of highly eccentric exoplanets.

\end{abstract}


%%%%%%%%%%%%%%%%%%%%%%%%%%%%%%%%%%%%%%%%%%%%%%%%%%

%%%%%%%%%%%%%%%%% BODY OF PAPER %%%%%%%%%%%%%%%%%%

\section{Introduction} \label{sec:intro}

After dominating the early days of exoplanetology, the radial velocity (RV) technique continues to play an outstanding role in exoplanet discovery. With observational baselines now extending over several decades, the RV technique is currently the main method for discovering long-period exoplanets \citep{Gregory10, Marmier13, Kane19, Rickman19}.

The radial velocity method does however have its limitations, of which the best known is the sin $i$ degeneracy that leaves the orbital inclination unknown. This constrains the determination of mass to a lower limit ($m\sin i$) and means that the true mass of RV planets can only be determined if the inclination can be constrained through other methods, which remains a challenging endeavour. The most promising way of measuring orbital inclinations is through astrometry, but thus far most astrometric data available have been insufficient to detect signals in the planetary regime. For example, \citet{Reffert11} used astrometry from the Hipparcos mission \citep{Hipparcos, HipparcosNew} to constrain the true masses of 310 substellar companion candidates but could confirm only nine as planets, all with upper limits on the masses rather than positive detections. The ongoing Gaia mission \citep{Gaia} is projected to dramatically improve this state of affairs by detecting many thousands of planets with astrometry \citep{Perryman14}, but this data will not be made public until Gaia Data Release 4, which is still years away.

A number of authors \citep[including, and not limited to,][]{Calissendorff18, Snellen18, Brandt19, Kervella19, Feng19} have pioneered the combination of Gaia and Hipparcos proper motion data to produce astrometry that can be used to detect companions on wide orbits. Although this amounts to only three measurements of tangential velocity (Hipparcos, Gaia, and the average motion between those observations), the high precision and long time-scale of these measurements means that, in combination with RV or imaging data, Hipparcos-Gaia astrometry allows for positive detection of planetary-mass companions in favourable cases \citep[e.g.][]{Dupuy19, DeRosa20, Xuan20, Li21}.

\citet{Blunt19} presented the discovery of HR~5183~b, a giant planet with a $74^{+43}_{-22}$ year orbital period discovered as a result of its spectacular periastron passage that occurred during 2017-2018. This remarkable companion has among the longest orbital periods of any planet discovered by the RV method, as well as one of the highest orbital eccentricities ($0.84 \pm 0.04$). Despite the extreme orbit of HR~5183~b, \citet{Kane19.HR5183} explored dynamical stability in the system and remarkably found that stable orbits in the Habitable Zone of the star are possible. 

HR~5183 has a probable stellar companion called HIP~67291 at a large sky separation of 490" ($\sim$15400~AU projected separation). The presence of this companion may have dynamically influenced the evolution of the planet HR~5183~b, a possibility that has recently been explored by \citet{Mustill22}. The authors find that orbital excitation of an initially circular planetary orbit by HIP~67291 is \boldnew{less likely than planet-planet scattering} to produce the observed eccentricity, but notably the combination of both of these excitation mechanisms results in the highest probability of reproducing the planetary orbit. As a corollary, \citet{Mustill22} interestingly find that the predicted distribution of mutual inclinations between the stellar and planetary orbits differ depending on the formation scenario, with planet-planet scattering inducing a broader distribution of mutual inclinations, potentially allowing for direct constraints on the planet formation history if the planet-binary mutual inclination can be measured precisely.

In this work we present the detection of HR~5183~b's astrometric reflex signal with Hipparcos-Gaia astrometry, and use this to provide the first constraints on the orbital inclination of this remarkable planet. We furthermore utilise precise Gaia astrometry of HR~5183 and HIP~67291 to explore the orbit of the binary, and combine the constraints on the inclinations of both orbits in an attempt to constrain the planet-binary mutual inclination. Finally, we demonstrate that there is a \bmaroon{moderately} significant excess of stellar multiples among systems containing highly eccentric planets, providing observational evidence that dynamical interactions with binary companions plays an important role in the origins of these extreme exoplanets.

In Section \ref{sec:method} we outline our method for modelling the planetary and stellar orbits. In Section \ref{sec:results} we document our results, followed by discussion in Section \ref{sec:discussion} and concluding remarks in Section \ref{sec:conclusions}.

\section{Method} \label{sec:method}

\begin{table*}
	\centering
	\caption{Parameters of the stellar binary used for the LOFTI model.}
	\label{table:binary_astrometry}
	\begin{tabular}{lcc}
		\hline
		Parameter & HR 5183 & HIP 67291 \\
		\hline
		Gaia EDR3 ID & 3721126409323324416 & 3721114933170707328 \\
		\hline
		Mass [${M_\odot}$] & $1.07 \pm 0.04$ $^a$ & $0.67 \pm 0.05$ $^a$ \\
		Parallax $\varpi$ [mas] & $31.7806 \pm 0.0257$ $^b$ & $31.8422 \pm 0.0157$ $^b$ \\
		RA proper motion $\mu_{\text{RA}}$ [\masyr{}] & \boldagain{$-510.20\pm0.05$} $^c$ & $-509.328 \pm 0.016$ $^b$ \\
		Declination proper motion $\mu_{\text{Dec}}$ [\masyr{}] & \boldagain{$-110.40\pm0.04$} $^c$ & $-110.975 \pm 0.011$ $^b$ \\
		Radial velocity [$\text{km}\,\text{s}^{-1}$] & \boldnew{$-30.66 \pm 0.10$} $^c$ & $-30.67 \pm 0.15$ $^b$ \\
		Projected separation [arcsec] & -- & $488.53692 \pm 0.00002$ $^b$ \\
		Position angle [degrees] & -- & $104.823966 \pm 0.000002$ $^b$ \\
		\hline
		\multicolumn{3}{l}{$^a$ From \citet{Blunt19} $^b$ From \citet{GaiaEDR3} $^c$ This work, after `de-projection' (see text).}\\
	\end{tabular}
\end{table*}

\subsection{Stellar parameters}

\citet{Blunt19} determined precise stellar parameters for HR~5183 and HIP~67291 using a combination of spectroscopic and photometric data plus Gaia DR2 parallaxes. Though updated parallaxes from Gaia EDR3 \citep{GaiaEDR3} are now available, the difference from the DR2 astrometry for these two stars is slight and we therefore do not expect to make significant improvements in parameter precision over \citet{Blunt19}. We therefore adopt the stellar parameters from that work, of which the most relevant for our purposes are the stellar masses, $1.07 \pm 0.04$ and $0.67 \pm 0.05$ $M_\odot$ respectively.

\subsection{The planetary orbit}

\subsubsection{Data}

In this work we use the same radial velocity dataset as \citet{Blunt19}, and we defer to that work for description of the RV data collection. The available RV data amounts to 175 observations with HJST/Tull, 104 at Lick/APF, and 78 with Keck/HIRES, of which 20 predate the 2004 spectrograph upgrade and 58 postdate it; as in \citet{Blunt19} we treat the pre- and post-upgrade HIRES radial velocities as separate datasets, allowing for a free offset between them.

For the astrometric data, we make use of the recently updated Hipparcos-Gaia Catalog of Accelerations \citep[HGCA;][]{Brandt21} based on proper motions from Gaia~EDR3 \citep{GaiaEDR3}. The data provided by the HGCA consists of three measurements of proper motion, in two co-ordinates each \citep{Brandt19}. These are the Hipparcos proper motion (\boldnew{$\mu_\text{H}$}) measured at approximately epoch 1991.25, the Gaia EDR3 proper motion (\boldnew{$\mu_\text{G}$}) measured at approximately 2016.0, and the mean Hipparcos-Gaia proper motion (\boldnew{$\mu_{\text{HG}}$}), derived from the change in sky position observed by the two telescopes. This last component has been referred to by a variety of \boldnew{names}; as in \citet{Venner21}, we refer to this measurement as the `Hip-Gaia proper motion'.

\subsubsection{Planetary orbit model} \label{subsec:planet_method}

The model used for the planetary orbit in this work is largely identical to the one in \citet{Venner21}, and we refer the reader to that work for detailed description of the techniques used here. In summary, we jointly fit the radial velocity and astrometric data using a two-body Keplerian model. This involves a total of 19 parameters, of which two are assigned Gaussian priors (the stellar mass $M_*$ and the parallax $\varpi$, for which we adopt the Gaia EDR3 parallax for HR~5183), seven describe the orbit of HR~5183~b (the orbital period $P$, the RV semi-amplitude $K$, the eccentricity $e$ and argument of periastron $\omega_1$ parameterised as $\sqrt{e}\sin\omega_1$ and $\sqrt{e}\cos\omega_1$, the time of periastron $T_p$, the orbital inclination $i$, and the longitude of node $\Omega$), two describe the proper motion of the system barycentre ($\mu_{\text{bary,RA}}$, $\mu_{\text{bary,Dec}}$), and the remaining eight are normalisation offsets and jitter parameters for each radial velocity dataset. The proper motions from Hipparcos and Gaia, which are effectively averaged over the respective $\sim$3.36 year and $\sim$2.76 year observation time-scales for those mission \citep{Hipparcos, GaiaEDR3astrometry}, are resampled based on their underlying observation times.\footnote{As the Gaia observation times are \bmaroon{not} yet available, we use the Gaia Observation Forecast Tool (\url{https://gaia.esac.esa.int/gost/}) to derive predicted measurement epochs.}

We use the Markov Chain Monte Carlo (MCMC) ensemble sampler \texttt{emcee~v3.0.2} \citep{emcee} to explore our parameter space. A total of 50 walkers were used to sample the 19-parameter model over $6\times10^{5}$ steps, with confirmation of convergence in the MCMC performed in the same way as in \citet{Venner21}. To derive our posterior samples, we then discarded 33\% of the chain as burn-in and saved every hundredth step from the 50 walkers. Using these posteriors, we then extracted the $68.3\%$ confidence intervals for the model parameters.

The main area of difference in our model with respect to \citet{Venner21} is that we adopt the same informed prior on the orbital period as \citet{Blunt19}:

\begin{equation}
\label{equation:Pprob}
p(P,t_d,B)=
\begin{cases}
1          & \text{if } (P-t_d)<B \\
(B+t_d)/P  & \text{otherwise}
\end{cases}
\:,
\end{equation}

where $P$ is the orbital period, $t_d$ is the duration of the periastron passage, and $B$ is the observational baseline. This is adapted from period priors used for long-period transiting exoplanets \citep{Vanderburg16, Kipping18}, and is motivated in principle because most information of the planetary properties is contained in the single observed periastron passage, an event which has a much shorter duration than the time-scale of observations; thus, the probability of detecting the planet can be approximated by the probability of observing its periastron passage. \citet{Blunt19} experimented with different values for $t_d$, a parameter which is difficult to define quantitatively, but found that their fits were indistinguishable and adopted $t_d=0$ for simplicity; we likewise follow this here.

The inclusion of Hipparcos-Gaia astrometry technically extends our observational baseline by approximately $\sim$2000 days over the radial velocities alone. However, this extension is due to the Hipparcos measurement, which has such large uncertainties it is not itself sensitive enough to detect the planetary signal, even if it were to observe the high-amplitude periastron passage (see Section \ref{subsec:planet_results}). As a result the the Hipparcos proper motion measurement does not, in practice, improve our capability to detect the reflex signal of HR~5183~b further back in time. Excluding this measurement from consideration for this purpose we therefore have a observational baseline equivalent to that from the RV data alone, and therefore our adopted observational baseline $B$ is identical to that of \citet{Blunt19}.

\subsection{The stellar binary}  \label{subsec:binary_method}

The possibility of physical association between HR~5183 and HIP~67291 has been posited several times in the literature, and this hypothesis was reviewed in detail by \citet{Blunt19}. Those authors were ultimately somewhat agnostic on whether or not the pair form a physical binary (although \citealt{Mustill22} have more recently reinforced the argument in favour of physical association)\bmaroon{;} we \bmaroon{therefore} aim to fully re-evaluate this hypothesis in this study.

It is first important to demonstrate that this pair is not a chance alignment of unrelated stars. To do this, we empirically infer the chance alignment probability based on the astrometric properties of stars near to the sky position of HR~5183 by querying Gaia EDR3 for sources within 0.5 degrees of HR~5183. We find that the probability of observing a random star with a parallax matching HR~5183's within $\pm$ 0.5 mas is $6.4\times10^{-4}$, and when adding a further constraint that the proper motions must match within $\pm$ 1 \masyr, the probability of chance alignment drops to $2.0\times10^{-5}$.  We show the results of this query in Appendix \ref{appendix:figures}. \bmaroon{Recently published results similarly support the association of the two stars;} in their catalogue of resolved binaries in Gaia EDR3 \citet{ElBadry21} estimate an even lower chance alignment probability of $1.91\times10^{-6}$ for this pair based on comparison between observed binaries and a synthetic catalogue of false alignments, \bmaroon{and the recent work of \citet{Kervella22} also supports a high probability of association for this pair.}
We therefore reject the hypothesis that HR~5183 and HIP~67291 are unrelated stars that happen to have similar space velocities and are coincidentally passing near to each other. However, this does not establish that the two stars are presently gravitationally bound, a question we will revisit in later sections of this work.

\subsubsection{Binary orbit model}

Accepting the hypothesis that HR~5183 and HIP~67291 are physically related, we next attempt to determine the orbital parameters of the prospective binary\boldnew{. For this purpose we use} LOFTI \citep[\texttt{lofti\_gaia},][]{Pearce20}\boldnew{, a code for binary orbit fitting which uses high precision Gaia astrometric data} as effectively instantaneous measurements of orbital position and velocity of stars in wide binaries. \boldnew{LOFTI} uses the Orbits For The Impatient \boldnew{rejection sampling algorithm} \citep[OFTI;][]{Blunt17} to \boldnew{efficiently} explore the allowed orbital parameter space\boldnew{, allowing for informative constraints on the orbits of wide binaries such as HR~5183-HIP~67291.}

\boldnew{To determine whether} this system is appropriate for modeling with LOFTI, it must first be \boldnew{demonstrated that} the Gaia astrometry is sufficiently precise and accurate for the task. For Gaia sources the most powerful data quality indicator is the Renormalised Unit Weight Error \citep[RUWE;][]{Lindegren18, GaiaEDR3astrometry}, a one-dimensional parameter derived from the astrometric $\chi^2$ normalised such that RUWE $=1.0$ reflects a robust fit to the astrometry, while a high RUWE value (e.g. RUWE $>1.4$ as advocated by \citealt{Lindegren18}, \bmaroon{though the results of studies such as \citealt{Belokurov20} and \citealt{Stassun21} suggest that RUWE values even slightly above $1.0$ can constitute evidence of unresolved companions}) indicates the fit to the Gaia astrometry is in some way subpar, such as for an unresolved close binary or due to significantly non-linear proper motion. In Gaia EDR3, HR~5183 and HIP~67291 have RUWEs of 1.108 and 0.982 respectively, which indicates that the fit to the Gaia EDR3 astrometry is of good quality for both stars and justifies our use of the Gaia astrometric data for modelling the orbit of this wide pair.

\boldnew{The formalism adopted by LOFTI assumes that the projected positions and velocities of the binary lie on a Cartesian plane. However, this assumption breaks down for large binary separations as a result of the curvature of the celestial sphere. Since the observable sky motions of the stars are projections of their true space velocities onto the celestial sphere, for widely separated stars the spherical nature of that surface means that their space velocities are projected with slightly differing vectors. This effect was explored by \citet{ElBadry19}, who showed that projection effects can cause two stars with identical space velocities to have significantly differing proper motions and radial velocities from the perspective of the observer. The author further} explored the significance of projection effects \boldnew{against binary separation} and found that these become non-negligible \boldnew{at} separations above $\gtrsim$0.1 pc for binaries within 120 pc\boldnew{. This} is comparable to the separation between HR~5183 and HIP~67291 \boldnew{(490" sky separation = 0.075~pc projected separation), so it is therefore necessary to account for projection effects in the velocity data when modelling the orbit of the binary.}

To correct for projection effects \boldnew{in the binary velocities we employ an approach of empirical `de-projection', which was previously used} in \citet{Venner21}. \boldnew{This works} by first converting the observed sky position and velocities of \boldnew{HR~5183} into its space velocities, and then converting these into the \boldnew{sky velocities} that would \boldnew{occur if the star instead} lay at the sky position of \boldnew{HIP~67291. This produces a set of hypothetical proper motions and radial velocities for HR~5183 that lie in the same plane as the observed values for HIP~67291, correcting for perspective effects arising from projection onto the celestial sphere and allowing for the direct comparison of their sky velocities.}%\footnote{\boldnew{We note that the choice of which component to apply this `de-projection' process to is arbitrary so long as the radial velocities of both components are known.}}

To do this we must first define \boldnew{the} astrometry and radial velocities for both stars. For HIP~67291 we adopt the parallax, proper motion, and radial velocity from Gaia EDR3 \citep{GaiaEDR3}. We note that, as a Hipparcos star, HIP~67291 is present in the HGCA \boldnew{\citep{Brandt21}}; however \boldnew{for this star the Hip-Gaia average proper motion is less precise than} the Gaia EDR3 proper motion, so we continue to use the EDR3 proper motion for this star. Notably \boldnew{all proper motion measurements of HIP~67291 in the HGCA are consistent with constant motion}, suggesting that the star does not have massive interior companions and therefore justifying the use of its astrometry to constrain the orbit of the wide binary. For HR~5183 we adopt the Gaia EDR3 parallax, however instead of the EDR3 proper motion we use the barycentric proper motion of (\boldnew{$-510.64 \pm 0.05$}, $-110.41 \pm 0.04$) \masyr{} derived from our model of the planetary orbit (see Section \ref{subsec:planet_results}), as this has been corrected for the perturbatory effects of HR~5183~b. For the adopted radial velocity we make use of the fact that HR~5183 is a Gaia RV standard star and was used as a validator for the Gaia DR2 radial velocities by \citet{Soubiran18}. Because the Gaia RV zero-point is calibrated to the SOPHIE RVs used in that work, the precise radial velocity of HR~5183 is directly comparable to the Gaia RV of HIP~67291 without instrumental offsets. Subtracting our best-fitting planetary orbit from the SOPHIE radial velocities provided by \citet{Soubiran18} we measure an absolute RV of $-30.50~\text{km}\,\text{s}^{-1}$ for HR~5183; we adopt an inflated uncertainty of $\pm 0.1~\text{km}\,\text{s}^{-1}$ for this value to accommodate for any systematics relating to the absolute RV reference frame.

Taking \boldnew{our} adopted astrometric and radial velocity data for HR~5183, we calculate stellar space velocities \boldnew{of \boldagain{$(U,V,W)=(-60.58\pm0.06, -55.07\pm0.05, -17.36\pm0.09)$}~km~s$^{-1}$ using \boldagain{the} equations of \citet{Johnson87}. By converting these to the sky velocities that would result at the position of HIP~67291 we \boldagain{obtain} a `de-projected' proper motion of \boldagain{($-510.20\pm0.05, -110.40\pm0.04$)}~\masyr{} and a radial velocity of $-30.66\pm0.10$~km~s$^{-1}$ for HR~5183.\footnote{\boldagain{We make an implementation of our `de-projection' technique available at \url{https://colab.research.google.com/drive/10a8URblzfdStzxNdCAb-Wnep_DV_-txj}.}} The perspective-corrected proper motion of the star differs by \boldagain{($+0.44, +0.01$)}~\masyr{} from our input values, a statistically significant correction which demonstrates the importance of accounting for projection effects in this system.}

\boldnew{We summarise our adopted system data in Table \ref{table:binary_astrometry}. Using these parameters as inputs for LOFTI, we ran the orbital model as described in \citet{Pearce20} for a total of $5\times10^{5}$ samples.}

\subsubsection{Binary bound probability} \label{subsubsec:bound_prob}

Having revised the astrometric and radial velocity data for the binary, we are now in a position to reappraise the calculation of binding probability for the binary as conducted by \citet{Blunt19}. Briefly, we performed a Monte Carlo simulation of orbital velocities drawn from the `de-projected' proper motions and radial velocities, and determined the fraction of total velocity vectors that exceed the escape velocity at the current binary separation (v$_{\rm{esc}} = 0.37$ km~s$^{-1}$ for a test particle at 15400 AU separation from a central mass of 1.74 M$_\odot$). We report on the results of this simulation in Section \ref{subsec:binary_results}.

\section{Results} \label{sec:results}

\subsection{HR 5183 b} \label{subsec:planet_results}

\begin{figure}
	\includegraphics[width=\columnwidth]{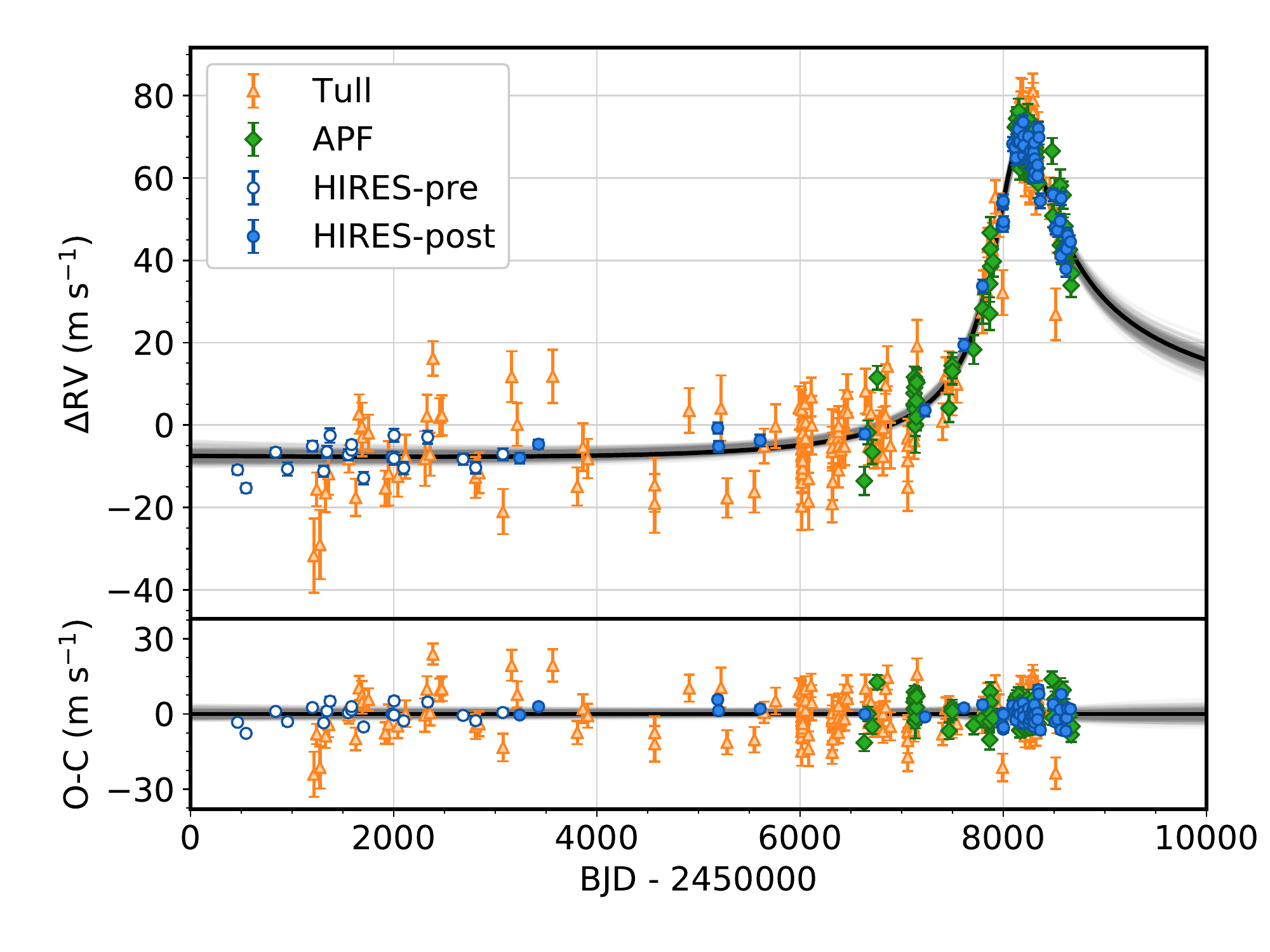}
	\caption{Our radial velocity model (top) and residuals (bottom) for HR~5183, showing the reflex signal of its planetary companion. The black line corresponds to the best-fitting model, while the grey lines are drawn randomly from the posterior distribution and demonstrate the range of plausible models. Our fit to the radial velocities is essentially identical to that of \citet{Blunt19}.}
	\label{figure:planet_RV}
\end{figure}

\begin{table*}
	\centering
	\caption{Parameters of HR 5183 b. All values are medians and 1$\sigma$ confidence intervals from the posterior distributions.}
	\label{table:planet_orbit}
	\begin{tabular}{lcc}
		\hline
		Parameter & \citet{Blunt19} & This work \\
		\hline
		Period $P$ [years] & $74^{+43}_{-22}$ & $102^{+84}_{-34}$ \\
		Period $P$ [days] & $\left(27000^{+16000}_{-8000}\right)$ & $37200^{+30700}_{-12400}$ \\
		RV semi-amplitude $K$ [\ms{}] & $38.25^{+0.58}_{-0.55}$ & $38.4 \pm 0.6$ \\
		Eccentricity $e$ & $0.84 \pm 0.04$ & $0.87 \pm 0.04$ \\
		Argument of periastron (primary) $\omega_1$ [degrees] & $\left(340 \pm 2\right)$ & $339.9 \pm 1.8$ \\
		Time of periastron $T_0$ [JD] & $2458121 \pm 12$ & $2458122 \pm 12$ \\
		Secondary minimum mass $m_2\sin i$ [$M_J$] & $3.23^{+0.15}_{-0.14}$ & $3.24 \pm 0.12$ \\
		Relative semi-major axis $a$ [AU] & $18^{+6}_{-4}$ & $22.3^{+11.0}_{-5.3}$ \\
		Periastron distance $a(1-e)$ [AU] & $2.88^{+0.09}_{-0.08}$ & $2.87^{+0.08}_{-0.08}$ \\
		Apoastron distance $a(1+e)$ [AU] & -- & $41.8^{+22.1}_{-10.6}$ \\
		\hline
		Orbital inclination $i$ [degrees] & -- & $89.9^{+13.3}_{-13.5}$ \\
		Longitude of node $\Omega$ [degrees] & -- & $224.0^{+18.2}_{-20.3}$ \\
		$\sin i$ & -- & $0.987^{+0.012}_{-0.039}$ \\
		Orbital velocity semi-amplitude $\frac{K}{\sin i}$ [\ms{}] & -- & $39.1^{+1.6}_{-0.9}$ \\
		Secondary mass $m_2$ [$M_J$] & -- & $3.31^{+0.18}_{-0.14}$ \\
		\hline
		Barycentric RA proper motion $\mu_{\text{bary,RA}}$ [\masyr{}] & -- & \boldnew{$-510.64 \pm 0.05$} \\
		Barycentric declination proper motion $\mu_{\text{bary,Dec}}$ [\masyr{}] & -- & $-110.41 \pm 0.04$ \\
		\hline
		Tull RV offset $\gamma_{\text{T}}$ [\ms{}] & $-19.2^{+1.9}_{-2.1}$ & $-20.6^{+2.0}_{-2.2}$ \\
		APF RV offset $\gamma_{\text{APF}}$ [\ms{}] & $-47.2^{+2.0}_{-2.2}$ & $-48.6^{+2.1}_{-2.4}$ \\
		HIRES pre-upgrade RV offset $\gamma_{\text{H,pre}}$ [\ms{}] & $-52.6^{+1.3}_{-1.5}$ & $-53.6^{+1.5}_{-1.7}$ \\
		HIRES post-upgrade RV offset $\gamma_{\text{H,post}}$ [\ms{}] & $-52.4^{+2.0}_{-2.1}$ & $-53.8^{+2.1}_{-2.3}$ \\
		\hline
		Tull RV jitter $\sigma_{\text{T}}$ [\ms{}] & $5.8^{+0.6}_{-0.5}$ & $5.8^{+0.6}_{-0.5}$ \\
		APF RV jitter $\sigma_{\text{APF}}$ [\ms{}] & $3.7^{+0.5}_{-0.4}$ & $3.7^{+0.5}_{-0.4}$ \\
		HIRES pre-upgrade RV jitter $\sigma_{\text{H,pre}}$ [\ms{}] & $3.4^{+0.8}_{-0.6}$ & $3.3^{+0.8}_{-0.6}$ \\
		HIRES post-upgrade RV jitter $\sigma_{\text{H,post}}$ [\ms{}] & $3.3 \pm 0.4$ & $3.3 \pm 0.4$ \\
		\hline
	\end{tabular}
\end{table*}

\begin{figure*}
	\includegraphics[width=\columnwidth]{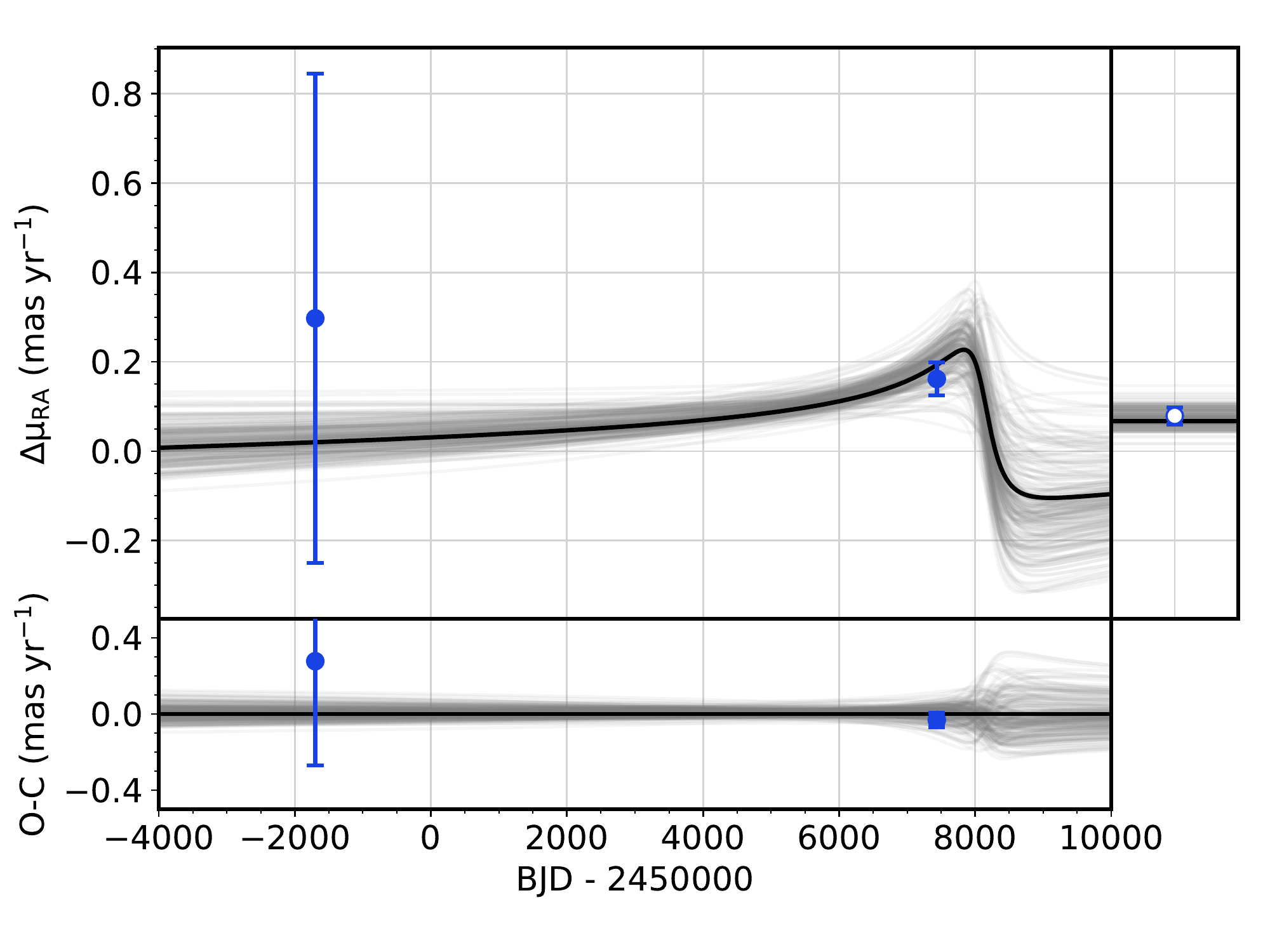}
	\includegraphics[width=\columnwidth]{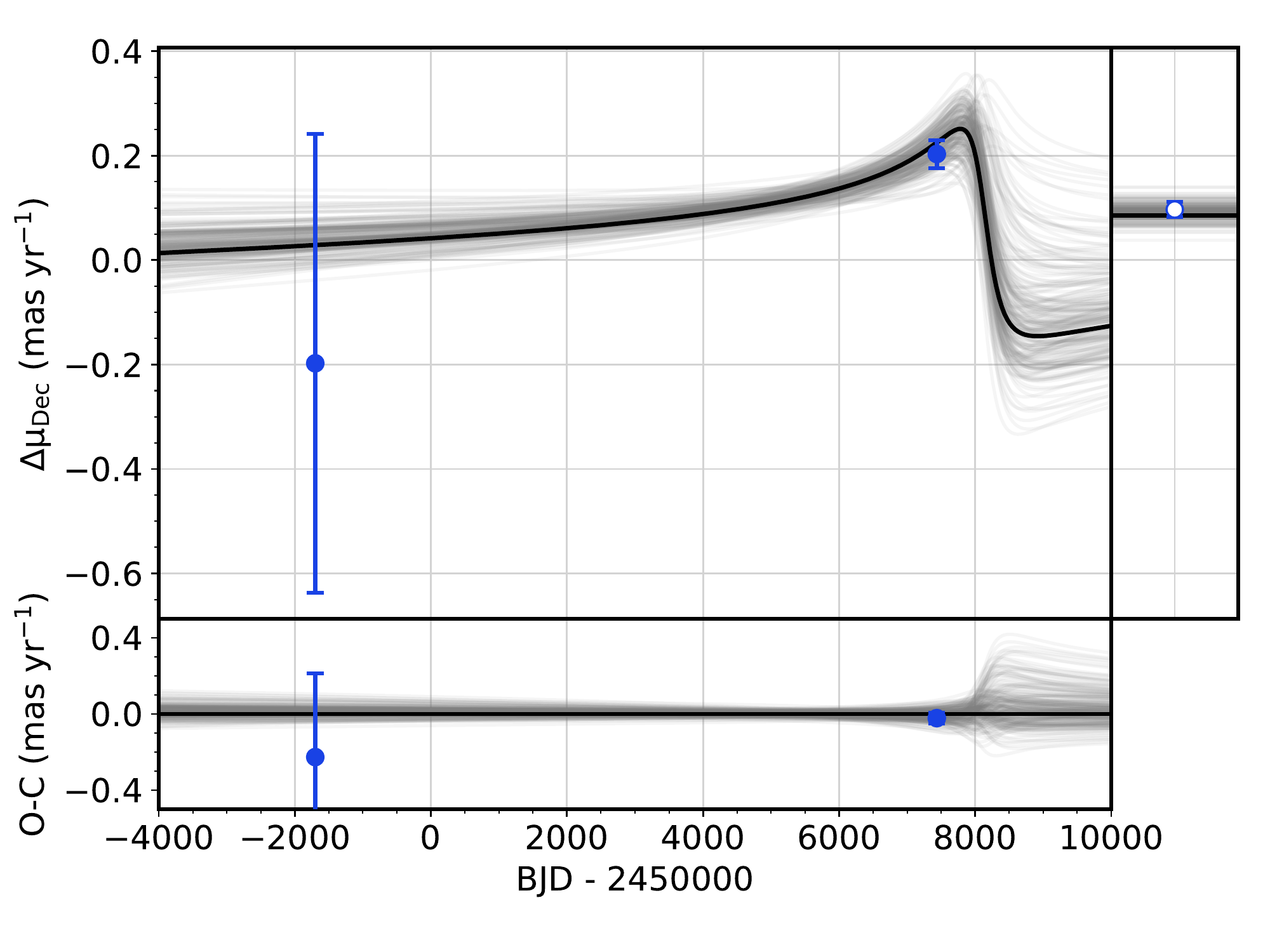}
	\caption{Our model for the proper motion of HR~5183, normalised to the star-planet barycentre, in right ascension (left) and declination (right). The two filled points in the main panels are the Hipparcos and Gaia proper motions while the unfilled points in the side panels represent the Hip-Gaia proper motions, which are averaged over the interval between the Hipparcos and Gaia observations. Due to the relatively large uncertainty on the Hipparcos proper motion almost all of our astrometric constraints are derived from the Gaia and Hip-Gaia measurements, which together imply an edge-on orbital inclination of $89.9^{+13.3}_{-13.5}\degree$ for HR~5183~b.}
	\label{figure:planet_PM}
\end{figure*}

The results of our orbital model for HR~5183~b are \boldnew{provided} in Table \ref{table:planet_orbit}. We measure a planetary orbital inclination of $89.9^{+13.3}_{-13.5}\degree$ and a longitude of node of $224.0^{+18.2}_{-20.3}\degree$. Our value for the orbital inclination is consistent with an edge-on orbit for the planet, and as a result our true mass of $3.31^{+0.18}_{-0.14}$~$M_J$ for HR~5183~b is almost identical to its minimum mass.

Our fit to the RV data is shown in Figure~\ref{figure:planet_RV}, and it is visually very similar to the RV model of \citet{Blunt19}. In comparison to that work we find a slightly longer orbital period and higher eccentricity and semi-major axis, although these are consistent within the uncertainties; all other parameters that can be compared are essentially identical. To test whether these differences are a result of the inclusion of astrometric data we additionally ran a RV-only version of our model, and found that this model indeed resulted in identical results as \citet{Blunt19}. Why exactly the inclusion of astrometric data leads to a preference for slightly longer orbital periods for HR~5183~b is not immediately clear, but one possibility is that the shortest orbital periods allowed by the radial velocities would result in a proper motion anomaly between the Gaia and Hip-Gaia measurements that is too large to match the observed signal.

In Figure~\ref{figure:planet_PM} we display our model for the proper motion data. This demonstrates that the Hipparcos measurement is insensitive to the planetary signal, hence our constraints on the orbital inclination and longitude of node are largely derived from the difference between the Gaia and Hip-Gaia proper motions. It is also evident that the Gaia measurement is temporally fortuitously close to the maximum velocity displacement of HR~5183, allowing us to confidently detect the proper motion signal generated by HR~5183~b.

\begin{figure*}
	\includegraphics[width=\textwidth]{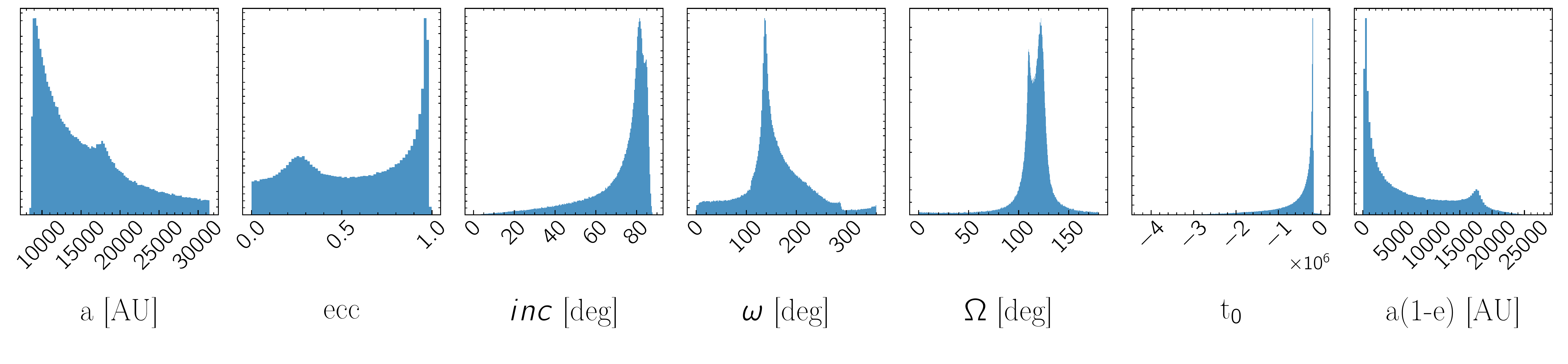}
	\caption{Posterior distributions of parameters from our LOFTI model of the HR~5183-HIP~67291 orbit. From left to right, the parameters are the semi-major axis, eccentricity, orbital inclination, argument of periastron, longitude of ascending node, time of periastron passage, and periastron distance. The long tails in semi-major axis and time of periastron have been truncated for clarity. Note that the longitude of node has been limited to the range (0, 180)$\degree$ since this parameter is degenerate over the full 360$\degree$ range.}
	\label{figure:binary_orbit_posteriors}
\end{figure*}

\begin{table*}
	\centering
	\caption{Orbital parameters for the HR~5183-HIP~67291 binary.}
	\label{table:binary_orbit}
	\begin{tabular}{lcccc}
		\hline
		Parameter & Median & Mode & 68\% CI & 95\% CI \\
		\hline
		Relative semi-major axis $a$ [AU] & \boldnew{$16400$} & \boldnew{$9000$} & \boldnew{$\left(8600, 22700\right)$} & \boldnew{$\left(8500, 67800\right)$} \\
		$\log P$ [years] & $6.20$ & \boldnew{$5.82$} & \boldnew{$\left(5.78, 6.42\right)$} & \boldnew{$\left(5.77, 7.13\right)$} \\
		Eccentricity $e$ & \boldnew{$0.59$} & \boldnew{$0.96$} & \boldnew{$\left(0.35, 0.98\right)$} & \boldnew{$\left(0.07, 0.98\right)$} \\
		Orbital inclination $i$ [degrees] & \boldnew{$79$} & \boldnew{$81$} & \boldnew{$\left(72, 86\right)$} & \boldnew{$\left(35, 87\right)$} \\
		Argument of periastron $\omega$ [degrees] & \boldnew{$148$} & \boldnew{$137$} & \boldnew{$\left(110, 217\right)$} & \boldnew{$\left(8, 288\right)$} \\
		Longitude of node $\Omega$ [degrees] & \boldnew{$118$} & \boldnew{$122$} & \boldnew{$\left(107, 128\right)$} & \boldnew{$\left(68, 165\right)$} \\
		Time of periastron $T_0$ [$10^{5}$ years \bmaroon{CE}] & \boldnew{$-4.83$} & \boldnew{$-2.00$} & \boldnew{$\left(-11.06, -1.72\right)$} & \boldnew{$\left(-63.82, 0.02\right)$} \\
		Periastron distance $a(1-e)$ [AU] & $6800$ & \boldnew{$270$} & \boldnew{$\left(130, 14900\right)$} & \boldnew{$\left(100, 54500\right)$} \\
		\hline
	\end{tabular}
\end{table*}

\subsection{Binary orbit} \label{subsec:binary_results}

Our parameter posteriors for the HR~5183-HIP~67291 binary orbit are shown in Figure \ref{figure:binary_orbit_posteriors} and listed in Table \ref{table:binary_orbit}. \boldnew{Uncertainties on the orbital parameters are generally large}, an explicable circumstance considering the large scale of the binary orbit, but we are \boldnew{nevertheless} able to derive meaningful constraints on all parameters. The semi-major axis \boldnew{posterior} displays a sharp peak at \boldnew{9000}~AU, with a long and essentially interminate tail towards larger values; \boldnew{54\%} of the posteriors have a semi-major axis \boldnew{higher} than the 15400~AU projected separation. The orbital eccentricity \boldnew{shows} a \boldnew{sharp peak at $e=0.96$ with a tail spanning all the way to to $e=0$, with a modest secondary peak at $e\approx0.25$}. As a result the periastron distance is less constrained than the semimajor axis, with a generally rising probability towards lower separations. \boldnew{The orbital period of} the binary is reasonably well-constrained \boldnew{in our model} with a modal value of \boldnew{$\sim7\times10^{5}$} years, approximately four orders of magnitude larger than the planetary orbital period.

Regarding the orbital inclination, our posterior \boldnew{peaks sharply at $i=81\degree$, followed by} a significant tail stretching all the way to $i=0\degree$. The longitude of node \boldnew{has a strong peak at $\Omega=122\degree$ along with a secondary peak at $\Omega=110\degree$, however, the entire} allowed parameter space is sampled in the posterior \boldnew{at low probability;} furthermore, since the radial velocities of the two stars are statistically indistinguishable after `de-projection' (see Table \ref{table:binary_astrometry}), we are unable to break the $180\degree$ degeneracy in the longitude of node. Thus, the posteriors in longitude of node are mirrored around $180\degree$, and the distribution given in Table \ref{table:binary_orbit} and Figure \ref{figure:binary_orbit_posteriors} are folded for clarity. \boldnew{This degeneracy likewise affects the argument of periastron, which shows a strong maximum at $\omega=137\degree$ for the $\Omega<180\degree$ orbital solution.} %Compared with the planetary argument of periastron ($\omega=339.9\pm1.8\degree$) the two orbits are likely to be aligned or anti-aligned in this parameter depending on the resolution of the nodal degeneracy, with $\Delta\omega\approx203\degree$ or $\approx23\degree$ respectively.

The significant probability \boldnew{density for orbital eccentricities approaching $e=1$ in our posterior} suggests that hyperbolic orbits are consistent with the available data \boldnew{for the binary}. We are unable to directly test this possibility in our model since LOFTI imposes an $e<1$ constraint on trial orbits, so we instead use the methodology outlined in Section \ref{subsubsec:bound_prob} to determine the fraction of \boldnew{non-hyperbolic orbits. From this model} we find a bound probability of \boldnew{73\%} for the binary, \boldnew{higher than the 44\% found by \citet{Blunt19} using a similar method}. \boldnew{Though this result is already encouraging, in Section \ref{subsec:binary_discussion} we further argue that this estimate is likely to be pessimistic and that the binary is almost certainly gravitationally bound.}

\subsection{Planet-binary mutual inclination}

\begin{figure*}
	\includegraphics[width=\columnwidth]{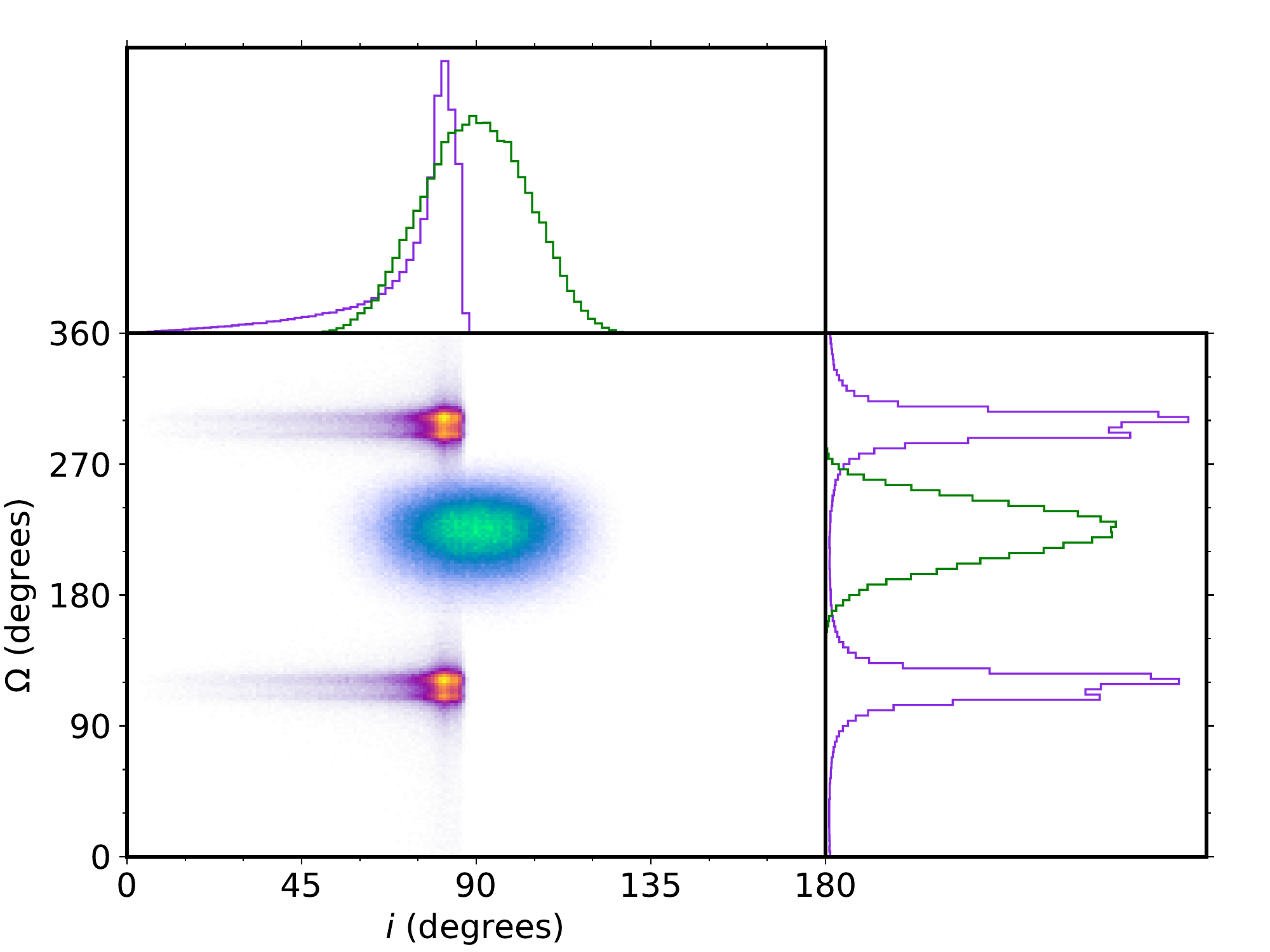}
	\includegraphics[width=\columnwidth]{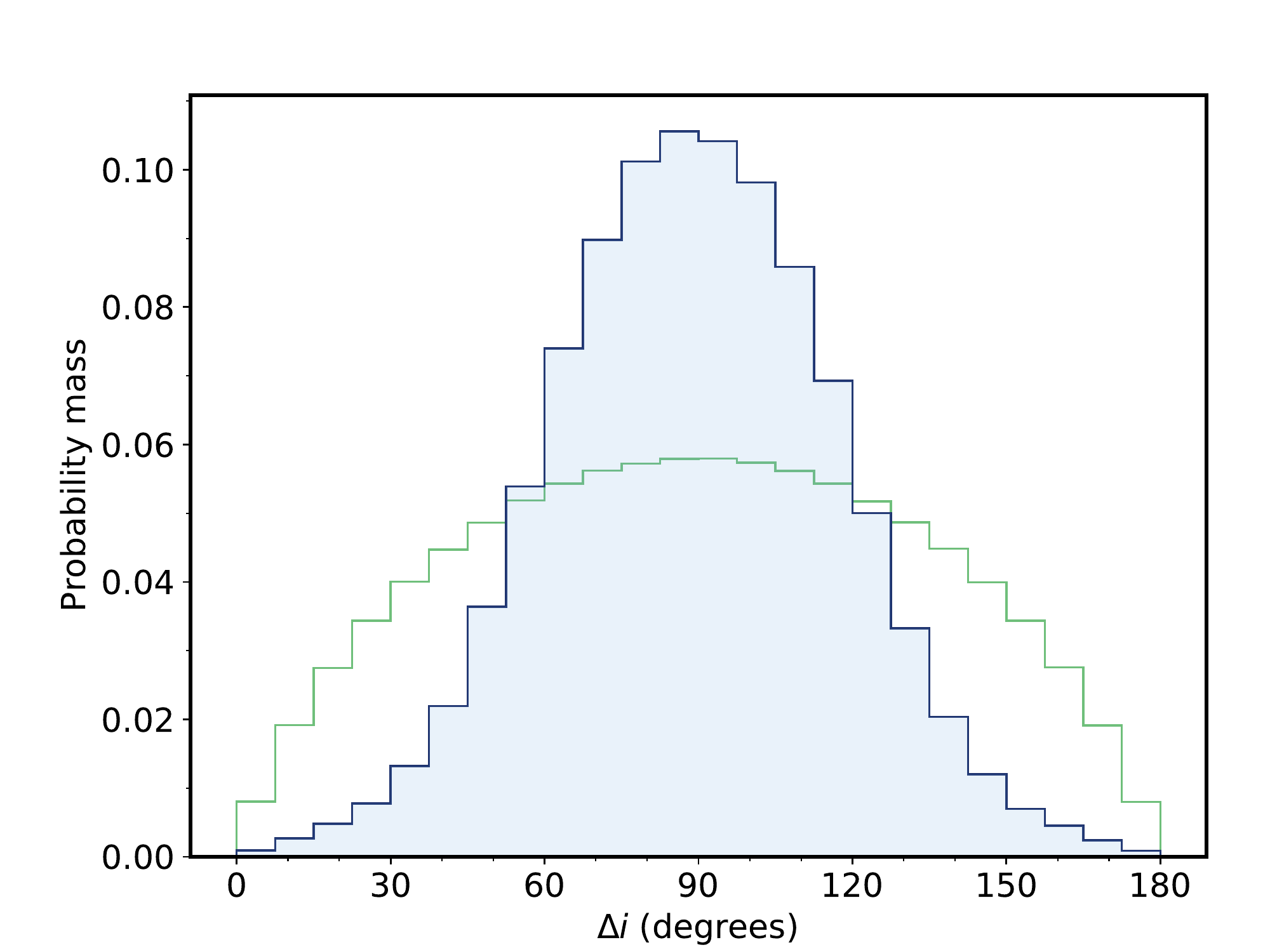}
	\caption{(Left) Distribution of inclinations and longitudes of node for HR~5183~b (blue-green) and HIP~67291 (purple-yellow). The planetary and stellar parameters appear to differ in their distributions \boldnew{for the longitude of node}, suggesting misalignment between the two orbits. (Right) Distribution of planet-binary mutual inclinations \boldnew{$\Delta i$}. The blue histogram reflects our model distribution while the green histogram represents \bmaroon{the prior, which is uniform in $\cos\Delta i$}. \boldnew{We} observe a preference for misaligned orbits \boldnew{($\Delta i\approx90\degree$), which remains even after acknowledging the bias towards misaligned orbits resulting from the non-uniform prior.}}
\label{figure:inclinations}
\end{figure*}

Having now constrained the planetary and stellar orbits, we plot their orbital inclinations and longitudes of node in \boldnew{the left panel of} Figure \ref{figure:inclinations}. \boldnew{The precision of the orbital parameters are generally similar in both cases, although the inclination distribution for HIP~67291 is significantly more asymmetrical than for HR~5183~b and there is a clear} $180\degree$ degeneracy in the longitude of node \boldnew{for the binary orbit.} The distributions of these parameters are suggestive of misalignment between the two companions; in particular, \boldnew{while the two orbital inclinations are consistent with alignment at $i\approx80\degree$, the location of the $225\degree$ peak in the planetary longitude differs significantly from the maxima at $120\degree$ and $300\degree$ for the binary orbit.} However, the parameter distributions for both companions are too broad to confidently infer orbital misalignment from this figure \boldnew{alone}, so it is necessary to quantitatively assess the mutual inclination between the two orbits.

For any two companions $b$ and $c$ orbiting a star, the mutual inclination \boldnew{$\Delta i$} between their two orbits can be defined as

\begin{equation}
\label{equation:mutual_inclination}
\cos\Delta i=\cos i_b \cos i_c + \sin i_b \sin i_c \cos(\Omega_b - \Omega_c)
\:,
\end{equation}

\citep{DeRosa20, Xuan20}. Since all of these parameters have been determined to some extent in our results, we are able to directly constrain the mutual inclination between HR~5183~b and \boldnew{HIP~67291}. \boldnew{However}, it is important to note that the resulting prior on the mutual inclination is uniform in $\cos\Delta i$ rather than in $\Delta i$, meaning that a randomly sampled posterior can have the appearance of supporting a misalignment.

Our resulting mutual inclination distribution is shown in the right panel of Figure \ref{figure:inclinations}. We compare our \bmaroon{model} distribution with \boldnew{an empirical} prior drawn from uniform sampling of the orbital inclinations and longitudes of node. Our distribution of mutual inclinations is \boldnew{indicative} of planet-binary orbital misalignment, \boldnew{with a peak in the mutual inclination of $\Delta i=90\degree$. This is primarily driven by the $\approx75\degree$ or $\approx105\degree$ disagreement in the longitudes of node, whereas the contribution from the orbital inclinations is generally small. However it is also evident that the prior is biased towards misalignment, with the peak in the uniformly sampled distribution likewise lying at $\Delta i=90\degree$. This means that the significance of our mutual inclination measurement must be overestimated. Nevertheless, the posterior distribution clearly favours mutual inclinations around $\approx90\degree$ more strongly than can be explained by prior bias alone; we find 3$\sigma$ lower limit of $\Delta i>13\degree$ for the mutual inclination, whereas the equivalent for the uniformly sampled distribution is just $>4\degree$. Arbitrarily setting up $\Delta i<10\degree$ as a definition for aligned orbits, we find that $1.33\%$ of the prior qualifies as aligned whereas only $0.16\%$ of the posterior distribution does the same, resulting in an odds ratio of $8.4$. These statistics strongly suggest that the misalignment between the planetary and stellar orbits is significant, even \boldagain{when considering the} bias of the prior towards such a result.}

%We therefore conclude that the planetary and stellar orbits are misaligned. However, while our posterior distribution can be visually distinguished from the prior, it is evident that our results must be biased towards mutual inclinations of $\Delta i\approx90\degree$. This means that the significance of the orbital mutual inclination must be overestimated, and we therefore do not regard the aforementioned posterior values to be definitive.} Improvements on the parameter constraints for both orbits are required before \boldnew{the} planet-binary mutual inclination can be \boldnew{quantified} with confidence.

%However, our mutual inclination distribution is \bmaroon{very similar to the prior, differing only in being slightly more centralised at $\Delta i=90\degree$}. This indicates that we are not able to significantly distinguish our planet-binary mutual inclinations from a random distribution based on our present results; thus, while our model suggests a mutual inclination between the planetary and stellar orbits, we consider this result to be tentative. Improvements on the parameter constraints for both orbits -- especially for the binary orbit, which presently has much broader uncertainties -- are required before a planet-binary mutual inclination can be detected with confidence.

\section{Discussion} \label{sec:discussion}

\subsection{HR 5183 b -- an eccentric giant planet with an edge-on orbit} \label{subsec:planet_discussion}

Our joint RV-astrometric model for HR~5183~b confirms the extremely high-eccentricity and long-period orbit for the planet found by \citet{Blunt19}. With its orbital inclination now measured as $89.9^{+13.3}_{-13.5}\degree$, we are able to unambiguously confirm that HR~5183~b is a planet with a true mass of $3.31^{+0.18}_{-0.14}$~$M_J$. The near-equivalence of the minimum and true planetary masses is relevant for our understanding of the dynamics of the HR~5183 system, since the dynamical studies of \citet{Kane19} and \citet{Mustill22} have both implicitly assumed that the planetary minimum mass of $3.23^{+0.15}_{-0.14}$~$M_J$ measured by \citet{Blunt19} is a good approximation of the true mass; our results demonstrate that this assumption is valid.

It is notable that in the Gaia DR2 version of the HGCA \citep{Brandt18}, astrometric acceleration of HR~5183 was not significantly detected. By contrast, in the Gaia EDR3 HGCA \citep{Brandt21} acceleration of HR~5183 was detected at $>$3$\sigma$ significance, allowing us to measure the orbital inclination of HR~5183~b. This rise in statistical significance can be attributed to two main factors. The first cause, more dominant of the two, is the substantial increase in astrometric precision between the two catalogues; \citet{Brandt21} estimates that the precision of the EDR3 HGCA reflects an improvement by a factor of $\sim$3 over the DR2 version, which \bmaroon{can itself} largely be ascribed to the significant improvement in astrometric precision of Gaia EDR3 over DR2 \citep{GaiaEDR3astrometry}. The second cause is the extended time coverage of Gaia EDR3. The timespan of astrometric observations used in the Gaia DR2 solution is 2014~Aug~22 -- 2016~May~23, while in Gaia EDR3 this has been extended to 2017~May~28 \citep{GaiaDR2astrometry, GaiaEDR3astrometry}. This time extension is highly relevant because, as is evident from Figure \ref{figure:planet_PM}, the average epoch of Gaia observations is \boldnew{only} $\sim$500 days prior to the maximum of the proper motion anomaly $\Delta\mu$ generated by HR~5183~b; as a result, the time extension of Gaia EDR3 has brought the average epoch of the Gaia astrometry closer to the epoch of maximum astrometric signal, increasing the detectability of HR~5183~b.

The nature of the available astrometric data does however constrain our ability to detect the reflex signal of HR~5183~b. The Hipparcos-Gaia proper motions used in this work are tangential velocities that are themselves time-averaged from position measurements taken over spans of several years. In contrast, the periastron passage of HR~5183~b is a rapid event, with a complete reversal in the sign of $\Delta\mu$ taking place in no more than a few hundred days; as a result the time-averaging of the proper motion data leads to a considerable loss of resolution around periastron passage, which necessarily limits our ability to precisely constrain the orbital inclination and longitude of node of HR~5183~b. We anticipate that once the full Gaia astrometric data is released with Gaia DR4 it will be possible to detect the planetary reflex signal with significantly finer time resolution, allowing for significantly reduced uncertainties for the orbital inclination and longitude of node for HR~5183~b compared to those measured here.

\subsubsection{Sky orbit and direct imaging} \label{subsec:direct_imaging}

\begin{figure}
	\includegraphics[width=0.85\columnwidth]{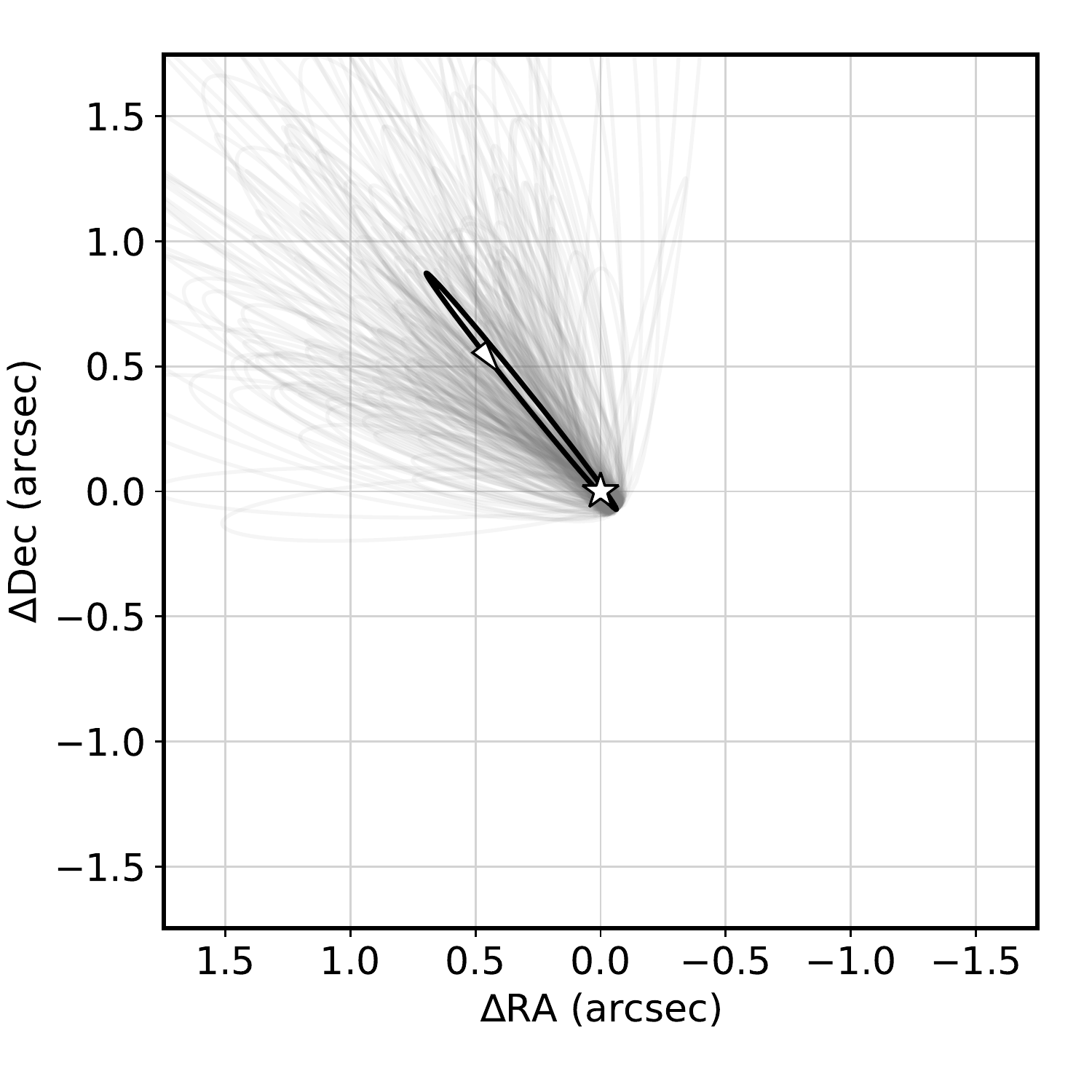}
	\caption{Our projected sky orbit for HR~5183~b. The star symbol at (0, 0) denotes the position of HR~5183. As in Figure~\ref{figure:planet_RV}, the black line corresponds to the best-fitting model while the grey lines are drawn randomly from the posteriors; the axes scales have been limited for clarity. The broad range of allowed orbits reflects the significant uncertainties in the orbital period, inclination, and longitude of node. The white arrowhead indicates the direction of motion for the best-fitting orbit.}
	\label{figure:planet_pos}
\end{figure}

Owing to our astrometric constraints on the orbital inclination and longitude of node of HR~5183~b we can predict the relative orbit of the planet around HR~5183, which we plot in Figure~\ref{figure:planet_pos}. The significant uncertainties in these parameters and the planetary orbital period manifest in the relatively wide range of sky orbits that are allowed by the available data, in particular the very broad distribution of possible separations at apoastron. Since the planetary orbit is observed edge-on the position angle of a given orbit is governed primarily by the longitude of node, hence the spread of orbits across a quarter of the figure is caused by the substantial $\sim$20$\degree$ uncertainty on this parameter.

\begin{figure}
	\includegraphics[width=\columnwidth]{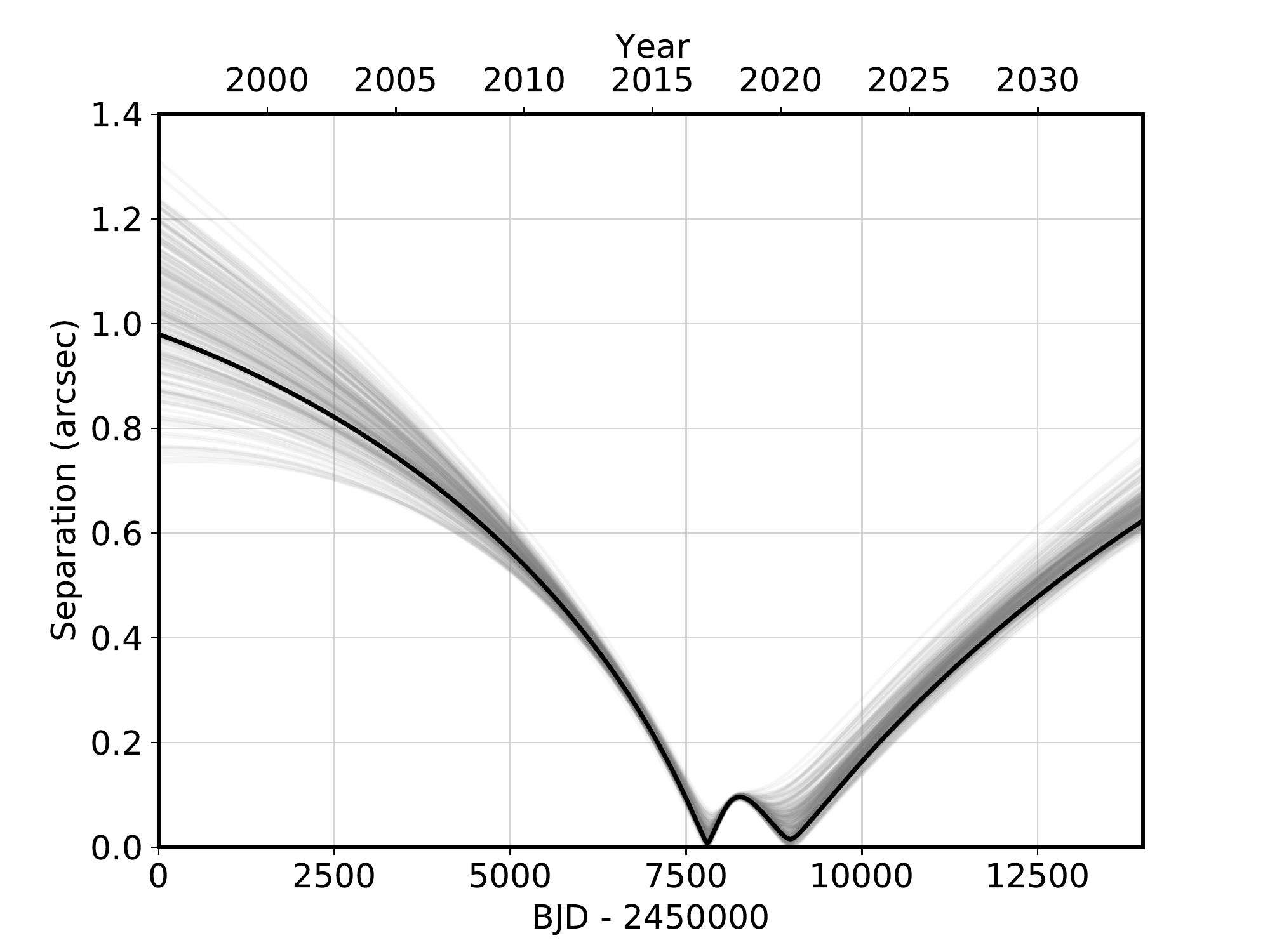}
	\caption{Planet-star sky separation over time for HR~5183~b. Minima in separation occur at inferior conjunction (BJD = 2457800) and superior conjunction (BJD$\approx$2459000); note that periastron counter-intuitively occurs close to the secondary peak in separation between these epochs.}
	\label{figure:planet_sep}
\end{figure}

Restricting ourselves to the planetary sky separation we find considerably tighter constraints than for the full sky orbit, as shown in Figure~\ref{figure:planet_sep}. Beginning with a wide spread of possible separations at earlier epochs, the planetary separation is well-constrained from BJD$\approx$2455000 up until inferior conjunction at BJD 2457800. Periastron counter-intuitively occurs at the secondary peak in separation at BJD 2458100, after which the planet passes through superior conjunction and then begins to separate from HR~5183. Subsequent to superior conjunction the planet-star separation remains well-constrained \bmaroon{with} the small degree of uncertainty primarily related to the orbital inclination\bmaroon{,} inclinations closer to 90$\degree$ corresponding to smaller separations.

Based on these results we can improve upon the observations made by \citet{Blunt19} regarding the prospects for direct detection of HR~5183~b. Assuming a random distribution of $i$ and $\Omega$, \citet[][\bmaroon{figures 7 -- 9}]{Blunt19} predicted the separation and contrast of HR~5183~b at a series of epochs in the range of 2020-2025. The separation distributions found by those authors are generally composed of a sharp peak, corresponding to edge-on orbital inclinations, followed by an extended tail towards wider separations caused by near-polar orbital inclinations. As we have measured an edge-on inclination for HR~5183~b we can exclude these low-probability tails, so our separations as plotted in Figure \ref{figure:planet_sep} correspond to the low-separation peaks in \citet{Blunt19}.

Because the detectability of the planet through direct imaging increases with increasing separation, and we have excluded the tails towards larger separations found in \citet{Blunt19}, the prospects for directly imaging HR~5183~b are restricted to the more challenging end of those presented in that work. Nevertheless, following the predicted contrasts given by those authors, the planet will reach the $300-500$ mas separation required for detection in the infrared by the second half of this decade for the entire range of plausible orbital inclinations \bmaroon{--} even despite the relatively old age and low mass of the planet. Thus, HR~5183~b is likely to be \bmaroon{detectable} with high-contrast imaging in the relatively near future. Direct \bmaroon{observation} of the planet will allow for study of its atmospheric properties, and additionally the precise astrometry provided by its detection would allow for further improvements in the measurement of the orbital parameters of HR~5183~b.

\subsubsection{Transit probability}

Our measurement of the planetary orbital inclination of $89.9^{+13.3}_{-13.5}\degree$ motivates us to revisit the possibility that HR~5183~b may, despite its long orbital period, be a transiting planet. As is well-known, a planet may transit its parent star if its orbit is observed sufficiently near to edge-on from our line of sight. The occurrence of transits can be parametrised by $b<1$, where $b$ is the impact parameter, a term defined as

\begin{equation}
\label{equation:impact_parameter}
b=\frac{a\cos i}{R_*} \left(\frac{1-e^2}{1-e\sin\omega}\right)
\:,
\end{equation}

Where $R_*$ is the stellar radius and all remaining parameters are as in Section \ref{subsec:planet_method} \citep[][\bmaroon{equation~7}]{Winn10}. Thus, the probability of observing transits is inversely proportional to the star-planet separation at inferior conjunction. Without knowledge of the planetary orbital inclination, \citet{Blunt19} inferred a transit probability $p_{\text{tra}}$ of $0.00185 \pm 0.00010$ for HR~5183~b assuming a uniform inclination distribution. Since our measurement of the planetary orbital inclination is fully consistent with an edge-on orbit, it is worth revisiting the transit probability based on the planetary orbital parameters measured in this work.

We adopt a stellar radius $R_*=1.53^{+0.06}_{-0.05}$ $R_\odot$ as in \citet{Blunt19}. First assuming a uniform distribution of orbital inclinations, we measure $p_{\text{tra}}=0.0010$. This is lower than found by \citet{Blunt19}, a difference which we ascribe to our preference for slightly longer orbital periods (see Section \ref{subsec:planet_results}). Secondly, when incorporating our constraints on the planetary orbital inclination we find $p_{\text{tra}}=0.0033$. The factor of $\sim$3 improvement in the transit probability appears to be quite modest considering that our inclination distribution peaks almost exactly at 90$\degree$, but seeing as the extremely narrow range of orbital inclinations where transits occur (approximately $\pm0.1\degree$) is two orders of magnitude smaller than our inclination uncertainty, the relatively small increase in transit probability is understandable. Nevertheless, it should be remembered that \bmaroon{if the measured} planetary orbital inclination \bmaroon{did differ} significantly from 90$\degree$\bmaroon{, it} would obviously greatly reduce the transit probability.

The previous inferior conjunction of HR~5183~b occurred in early 2017, and the next such event will only occur after another orbital period of the planet has elapsed. Due to the large uncertainties on this parameter we cannot provide a precise prediction of the next epoch of inferior conjunction, but our median orbital period of 102~years is sufficient to conclude that detection of planetary transits -- should they occur -- will not be possible in the near future. Nevertheless, future improvements on the planetary orbital parameters from radial velocity, astrometric, and direct imaging observations will allow for refined measurements of the probability and time of transit, and the conclusions made here will undoubtedly be revised long before the next inferior conjunction.

\subsection{HIP 67291 -- the wide stellar companion to HR 5183} \label{subsec:binary_discussion}

\begin{figure}
	\includegraphics[width=\columnwidth]{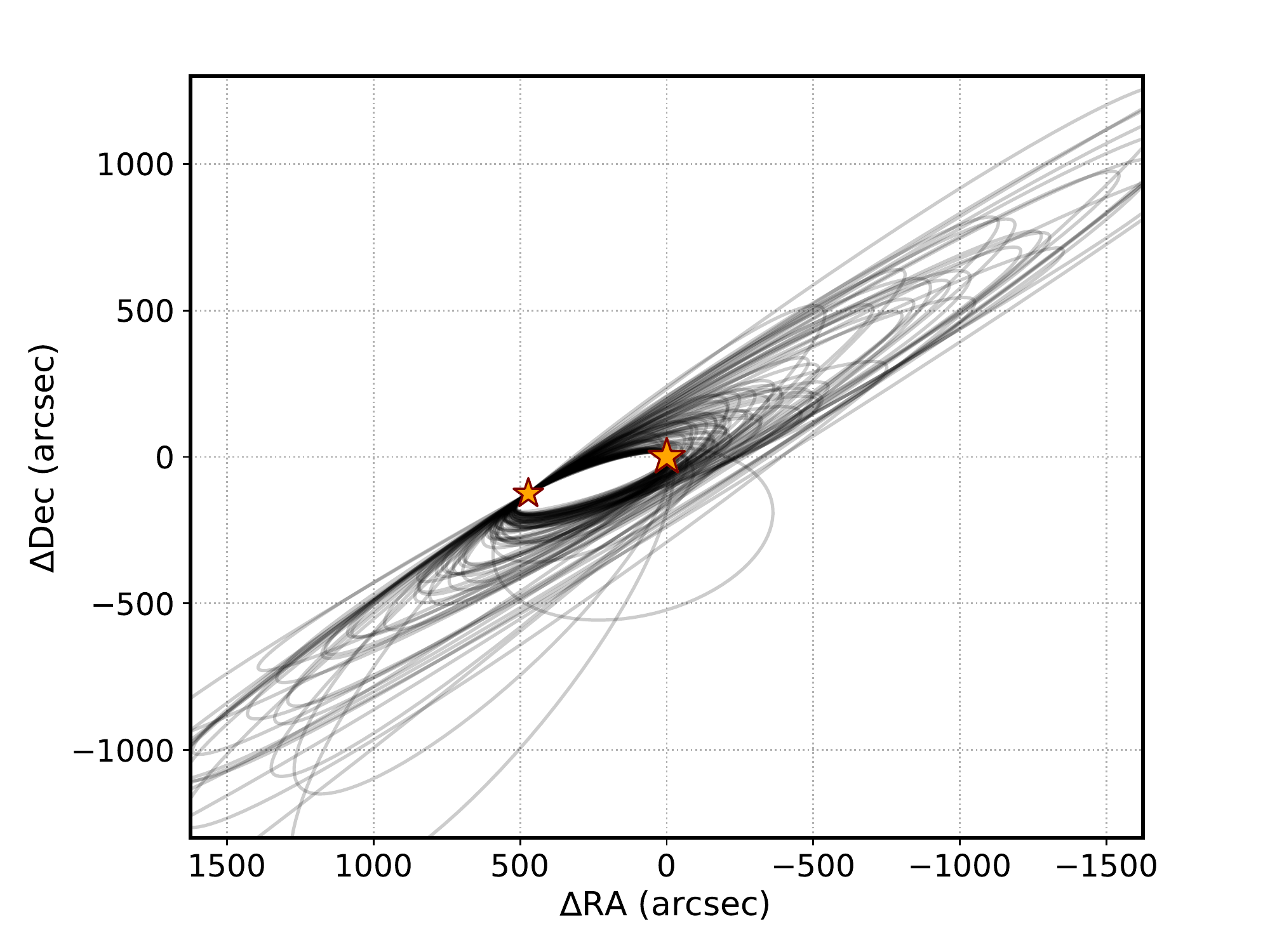}
	\caption{\boldnew{Our projected sky orbit for the stellar binary. The star symbols mark the positions of HR~5183 and HIP~67291 respectively, while the orbits are drawn randomly from the posteriors of the LOFTI fit. The direction of motion is clockwise for all orbits in the figure. There is a clear preference for near-edge on orbits ($i\approx80\degree$) for the binary.}}
	\label{figure:binary_orbit}
\end{figure}

In this work we have used LOFTI \citep{Pearce20} to constrain the orbital parameters of the wide HR~5183-HIP~67291 binary. With a sky separation of 490 arcseconds and a corresponding projected separation of 15400 AU, this system has one of the largest separations of any binary that has been modelled using this technique, and we were forced to account for perspective effects arising due to the non-linear nature of the celestial sphere in our model. We have found that the most likely binary orbits have semi-major axes of \boldnew{9000~AU, orbital inclinations of $81\degree$, and longitudes of node of $122\degree$ (with a $\pm180\degree$ degeneracy for the latter)}. However, all of these parameters have substantial uncertainties in our posteriors, and orbits with very different parameters should be considered plausible. \boldnew{We plot the sky orbit of the binary in Figure~\ref{figure:binary_orbit}, which shows the strong preference for highly inclined, near edge-on orbits for HR~5183-HIP~67291.}

\citet{Blunt19} performed a similar fit to the binary orbit as we have done, and it is informative to compare their model posteriors with ours. Our results differ in significant ways; \boldnew{for example,} their most probable bound orbits show a preference for semi-major axes around $\sim$25000~AU with very few orbits close to the peak in our results at \boldnew{9000~AU. Likewise,} their \boldnew{orbital inclination distribution is strongly peaked} at $\sim$100$\degree$ with a tail towards higher inclinations and seemingly no orbits below $i<90\degree$, \boldnew{while our inclination distribution peaks at $i=81\degree$ with none of our posteriors reaching above $i>90\degree$.} The clearest area of agreement between our results is in the \boldnew{orbital eccentricity,} where we both \boldnew{observe high values ($e>0.8$), and correspondingly in the periastron distance, where we find preferences for relatively small minimum separations ($a(1-e)<2000$~AU).}

Since the data underlying our fits are similar -- The difference between the Gaia DR2 solutions used by \citet{Blunt19} and the EDR3 solutions used in this work are generally insignificant -- we anticipate that these disparities are a result of methodological differences. First, we note that the astrometric acceleration generated by HR~5183~b substantially displaces the Gaia proper motion from the star-planet barycentre, and accounting for the planetary signal results in significant change to the stellar velocity. Secondly, our handling of perspective effects in the LOFTI model affects our velocities for HR~5183 as well, although the increased proper motion uncertainties arising from this procedure reduces the significance of these effects. Lastly, \citet{Blunt19} incorporated Gaia parallaxes of the two stars as positional information in their orbital model, which is not the case with LOFTI. While the Gaia DR2 parallaxes of HR~5183 and HIP~67291 imply a difference in stellar distances from the solar system of $-0.157 \pm 0.059$~pc (the negative sign meaning that HR~5183 is more distant), the Gaia EDR3 parallaxes for the pair result in a smaller difference of $-0.061 \pm 0.029$~pc. As a smaller distance between the two stars implies a larger escape velocity, it follows that the more similar stellar distances would entail lower orbital eccentricities given identical velocities. We anticipate that these differences can explain the dissimilarity between our binary orbit and that of \citet{Blunt19}.

To improve upon our constraints on the binary orbit, we suggest that stronger constraints on the stellar velocities are of importance. In particular, the radial velocity measurements for the two stars provide the largest portion of the uncertainties in the input data, and bringing these uncertainties below the 100 \ms{} level will be important for improving constraints on the binary orbit. We furthermore note that HIP~67291 has not previously been targeted for precise radial velocities to our knowledge; therefore, as well as contributing to improved precision on the binary orbital parameters, high-precision radial velocities of this star could reveal the presence of planets orbiting the secondary component of this wide pair. \bmaroon{One possible source of bias in the available data is that the Gaia~EDR3 astrometric solution for HR~5183 assumes a linear proper motion, which is demonstrably incorrect due to the periastron passage of HR~5183~b which occurs during the course of Gaia observations. Although we have corrected for the displacement in proper motion caused by HR~5183~b in our binary orbit model, it is possible that the underlying Gaia proper motion may be slightly biased due to the single-star nature of the EDR3 astrometric solution. Future Gaia data releases will account for non-linear proper motions in the astrometric model, thus accounting for this possible source of systematic error. Finally}, utilising the Gaia parallaxes for the binary orbit model may help to improve the constraints on some of the parameters.

\subsubsection{Appraising the binary bound orbit probability} \label{subsubsec:bound_prob_discussion}

In Section \ref{subsec:binary_results}, we found that a test particle model suggests that \boldnew{73\%} of possible orbits for HIP~67291 are bound (non-hyperbolic) based on the available data. \boldnew{This is a substantial increase from \citet{Blunt19}, who found a bound orbit fraction of 44\% based on a similar model. However, our result implies a considerable 27\% probability that HIP~67291 is gravitationally unbound, which justifies further consideration of the bound probability for the pair. In the following section we expand on the reasoning outlined by \citet{Blunt19} regarding the nature of HR~5183-HIP~67291 to argue that the pair is almost certainly a gravitationally bound binary.}

There are three possible hypotheses for the relationship between HR~5183 and HIP~67291:

\begin{enumerate}
	\item The two stars are unrelated, and their close physical separation is a coincidence.
	\item The two stars are physically related, and form a binary with a gravitationally bound orbit.
	\item The two stars are physically related, but are not presently on a bound orbit; they may have been bound in the past, but are now undergoing breakup.
\end{enumerate}

Hypothesis I was considered in Section \ref{subsec:binary_method}, and can be firmly rejected as improbable; \bmaroon{through a query of Gaia~EDR3 for stars with parallaxes matching within $\pm~0.5$~mas and proper motions within $\pm~1$~\masyr{} of the corresponding values for HR~5183,} we \bmaroon{empirically} estimate the probability of such a chance alignment in parallax and proper motion to be $2.0\times10^{-5}$. Hypothesis II corresponds to the \boldnew{73\%} fraction of bound orbits, while Hypothesis III reflects the remaining \boldnew{27\%} of unbound orbits. However, while Hypothesis II is a physically plausible arrangement for the two stars, Hypothesis III is not. HR~5183 and HIP~67291 would have to have become unbound relatively recently to conceivably be observed at their present separation; we can make a first-order estimate of this time-scale by taking the observed relative tangential velocity between the stars ($\approx$275 \ms{} from the proper motions in Table \ref{table:binary_astrometry}) and their projected separation of 15400 AU, which when put together result in a time-scale of separation of $\sim$2.6 $\times10^{5}$ years (disregarding gravitational effects, which would undoubtedly decrease this time-scale). Assuming that the probability of dissociation between the stars is uniform over the age of the system, and given an age for HR~5183 of $7.7^{+1.4}_{-1.2}$~Gyr \citep{Blunt19}, the probability of observing this pair in such a recent state of dissociation would be

\begin{equation}
\frac{\sim2.6\times 10^{5}}{\sim7.7\times 10^{9}}\approx 3.4 \times 10^{-5}
\:.
\end{equation}

This indicates that the probability of observing this system in a recently-unbound state is very low. With this informed prior in mind, it follows that Hypothesis III is far less probable than the naive \boldnew{27\%} estimate provided by the test particle model, and conversely the probability that the two stars are on a bound orbit must be much greater than \boldnew{73\%}. We thus conclude that HR~5183 and HIP~67291 form a physical, gravitationally bound system. Our investigation into the binary orbital parameters, and the planet-binary mutual inclination, can therefore be justified.

We anticipate that the prevalence of hyperbolic orbits in our test particle model reflect the proportionally large uncertainties of the stellar velocities as compared to the escape velocity of the system. If this is the case, future improvements in the observational measurement uncertainties may alter the estimation of the bound probability.

\subsubsection{Considerations on biases in the binary orbit}

Recently, \citet{FerrerChavez21} have explored biases in the fitting of orbits using the OFTI algorithm of \citet{Blunt17}; this forms the basis of the LOFTI model used to model the HR~5183-HIP~67291 orbit in this work, so their observations are likely to be relevant to our results. A key result of that work is that there is a strong degeneracy between orbital inclination and eccentricity, with edge-on circular orbits being difficult to distinguish from face-on eccentric orbits. This can perhaps be seen in our results since we find very broad posteriors for both of these parameters, but the peaks in these distributions suggest that orbits with near-edge on inclinations \textit{and} \bmaroon{high} eccentricities are generally preferred. However, those authors also observe that higher eccentricities leads to a bias towards edge-on inclinations, and it is quite possible that this bias is reflected in our results. Due to these biases, \citet{FerrerChavez21} find that in the case of near edge-on orbital inclinations the eccentricity will be largely unconstrained (which is presumably what can be seen in \boldnew{the breadth of our} eccentricity posterior) and that, qualitatively, the 68\% confidence interval for the eccentricity will recover the true eccentricity less than 68\% of the time.

Based on these observations it appears that we should avoid drawing much of significance from our binary eccentricity posterior, \boldnew{which shows a sharp peak at high values with a mode of $e=0.96$, as} it is highly likely that \boldnew{this} is statistically biased due to the various inherent degeneracies with the orbital inclination. It is less clear to what extent our binary orbital inclination should be trusted, but considering the magnitude of biases found for the eccentricity it should be considered distinctly possible that our inclination posteriors are biased away from the true value. If this is the case, it would have strong implications for the measurement of a mutual inclination between the planetary and binary orbits in this system\boldnew{; as our measurements of the orbital inclinations are both consistent with $i\approx90\degree$, any bias towards edge-on binary orbits would result in us \textit{underestimating} the mutual inclination in this system.}

\subsection{\boldnew{Planet-binary mutual inclination}}

In this study we have measured the orbital inclination and longitude of node of HR~5183~b for the first time, and we have combined these with the corresponding parameters for the orbit of HIP~67291 in an attempt to measure the planet-binary mutual inclination. Our results \boldnew{suggest a significant} preference for misaligned orbits, with \boldnew{a best-fit mutual inclination of $\Delta i\approx90\degree$ and a 3$\sigma$ lower limit of $\Delta i>13\degree$. This is primarily driven by disagreement between the longitudes of node of the planetary and binary orbits, which can be visually identified from the $\approx90\degree$ disagreement in the position angle of their sky orbits in Figure~\ref{figure:planet_pos} and Figure~\ref{figure:binary_orbit}. However we recognise that the prior for the mutual inclination is uniform in $\cos\Delta i$ rather than $\Delta i$, which means our results are biased towards the $\Delta i\approx90\degree$ that we observe in our posterior. Still, this bias is insufficient to explain our results alone, as the 3$\sigma$ lower limit in mutual inclination for a uniformly sampled distribution is just $\Delta i>4\degree$. Only $0.16\%$ of our posterior has a mutual inclination below $\Delta i<10\degree$ whereas the same statistic is $8.4$ times higher for the uniformly sampled distribution, a discrepancy which indicates that our preference for misaligned orbits is stronger than would be expected from chance. We therefore conclude that the orbits of the planetary and binary orbits are misaligned.}

\boldnew{In their recent study of the dynamical formation history of HR~5183~b, \citet{Mustill22} predicted that the mutual inclination between the planetary orbit and the orbit of the wide stellar companion HIP~67291 could potentially be used to distinguish between formation mechanisms for HR~5183~b. Their results suggest that very low or very high mutual inclinations ($\Delta i\approx0\degree$ or $\approx180\degree$) cannot be reproduced without invoking planet-planet scattering, while planet-binary interactions alone produce mutual inclinations of $\approx45\degree$ or $\approx135\degree$. Unfortunately, our measurement of the planet-binary mutual inclination is insufficiently precise} to test the predictions of \citet{Mustill22}\boldnew{; our empirical measurement of the planet-binary mutual inclination is consistent with a broad range of values, and is therefore consistent with formation of HR~5183~b with or without planet-planet scattering.}

While we are therefore unable to \boldnew{make inferences on the formation history of HR~5183~b} in this work, future studies may be able to do so if the uncertainties on the orbital inclinations and longitudes of node can be substantially reduced. For the planetary orbit, as noted in Section \ref{subsec:planet_discussion}, we anticipate that the full release of the Gaia astrometric data will allow for significantly more precise constraints on these parameters than we have been able to reach here; further in the future, detection of HR~5183~b with direct imaging will undoubtedly lead to even greater improvements in the measurement of the astrometric orbital parameters. However, \boldnew{for the HR~5183-HIP~67291 binary,} opportunities for improvement on the \boldnew{orbital} parameter precision are slimmer. As discussed in Section \ref{subsec:binary_discussion}, increased precision on the velocity measurements for the two stars is one of the most important areas for improvement for constraining the binary orbit, and while future Gaia releases will likely lead to more precise proper motion measurements, we suggest that observations with high-precision ground-based spectrographs will be important for improvement on the absolute radial velocities of HR~5183 and HIP~67291. Additionally, we have not used the individual parallaxes of the two stars to constrain the difference in their radial distances in our LOFTI model, and making use of this data may improve some of the constraints on the binary orbital parameters in future studies.

As discussed in the preceding section, \citet{FerrerChavez21} have observed significant biases in parameter determination from orbit fitting using the OFTI algorithm. In particular, the biases in inclination observed in that work may pose a significant challenge for measuring planet-binary mutual inclinations. Further investigation of these parameter biases is of importance before mutual inclinations in systems similar to the one studied in this work can be securely measured.

\subsubsection{Comparison with other studies}

In this section we highlight some studies that have explored mutual inclinations between planetary and stellar orbits in a similar way to our work, and comment on the implications of these results.

\citet{Li21} have recently used Hipparcos-Gaia astrometry to measure true masses for nine planets discovered through radial velocities using the \texttt{orvara} code \citep{orvara}. Two of these planets reside in binary systems (HD~106515~Ab and HD~196067~b), and like our work the authors used Gaia EDR3 astrometry to constrain the binary orbits. Unlike our results for HR~5183~b the planetary orbital inclinations are found to be bimodal, complicating the interpretation of mutual inclination in these systems. \bmaroon{For the HD~106515 system the results of \citet{Li21} would be consistent with alignment for the prograde orbital solution for HD~106515~Ab, but would suggest strong misalignment for their retrograde solution. In the HD~196067-196068 system, strong disagreement between the planetary and stellar orbital inclinations suggests misalignment.} It is notable that both of these planets have relatively large eccentricities (HD~106515~Ab $=0.571 \pm 0.012$, HD~196067~b $=0.70^{+0.14}_{-0.12}$); the possibility that the wide stellar companions in these systems have influenced the formation of these eccentric planets through dynamically interactions should be further investigated.

\citet{Newton19} presented the discovery of a transiting planet orbiting the primary component of the young visual binary DS Tucanae by the TESS mission \citep{TESS}, and as well as characterising the planetary companion the authors used LOFTI to constrain the orbit of the binary and determined the rotational inclination of DS Tucanae A. All three inclinations were found to be close to alignment in inclination, with a planetary orbital inclination of $89.5^{+0.34}_{-0.41}\degree$ (or $90.5^{+0.41}_{-0.34}\degree$), a stellar orbital inclination of $96.9 \pm 0.9\degree$, and a stellar rotational inclination of $82-98\degree$. Subsequent observations of the Rossiter-McLaughlin effect during transit have confirmed that the obliquity of the planetary orbit relative to the rotational axis of DS Tucanae A is low \citep{Montet20, Zhou20, Benatti21}. The general alignment of orbital planes and rotational axes in this system is suggestive of primordial coplanarity \citep{Montet20}.

\citet{Xuan20.inclinations} used Hipparcos-Gaia astrometry to measure the orbital inclinations of giant planets orbiting HD~113337 and HD~38529. Both systems contain two known planets, a debris disc, and widely separated M-type companions. The authors focused on measuring mutual inclinations between the planetary orbits and those of the debris discs, but also attempted to use LOFTI to constrain the orbit of HD~38529~B. While this fit provided only loose constraints on the stellar orbital parameters, this was sufficient for the authors to measure a planet-binary mutual inclination \boldnew{$\Delta i > 20\degree$} at 3$\sigma$ confidence. They further report a minimum planet-disc mutual inclination \boldnew{$\Delta i = 21-45\degree$} (1$\sigma$), which combine to suggest that the orbital planes in the HD~38529 system are generally misaligned.

%While these few results provide too little information to confidently formulate a synthesis at this stage, there is a hint of structure in the available data. The DS Tucanae system, with a binary semi-major axis of $157-174$~AU \citep{Newton19}, shows a remarkably strong alignment of orbital and rotational inclinations, with all measured inclinations in agreement within $\lesssim10\degree$. The HD 106515 system may be in a state of planet-binary orbital alignment as well - although the degeneracy in the planetary orbital inclination makes this interpretation uncertain - and here too the binary semi-major axis is relatively low, $335^{+96}_{-42}$~AU. In contrast, the HD 38529 and HD 196067-196068 systems appear to be inconsistent with planet-binary orbital alignment, and our results for the HR~5183-HIP~67291 system suggests that this may be the case here as well; all of these binaries have semi-major axes well beyond >1000~AU.

\bmaroon{From the foregoing results, it is notable that the examples of systems with smaller binary separations are more consistent with alignment (DS~Tucanae, HD~106515), while wider binaries show evidence for misalignment (HD~38529, HD~196067-196068, and perhaps HR~5183-HIP~67291 from this work).} In this context, the recent work of \bmaroon{\citet{Christian22}} lends significant support to the interpretation that binary semi-major axis plays an important role in planet-binary orbital alignment. In that work the authors used LOFTI to measure the orbital inclinations of a sample of visual binaries containing transiting planets and planet candidates discovered by the TESS mission and found a statistically significant excess of systems consistent with alignment between the planetary and stellar orbits, \bmaroon{further observing} that this overabundance of aligned systems is strongest for binary semi-major axes below <700~AU. Beyond >700~AU the distribution of binary orbital inclinations is more random, indicating that there is a significant fraction of systems with misaligned orbits. These observations satisfactorily match the properties of the systems discussed above. \bmaroon{\citet{Christian22}} suggest that a plausible explanation for this phenomenon is that during the protoplanetary disk phase, relatively close stellar companions on primordially misaligned orbits are capable of torquing the planet-forming disk into alignment prior to its dissipation. However, they note that for systems with semi-major axes below $\lesssim200$~AU it is possible that the binaries formed in primordial state of alignment through disk fragmentation, a hypothesis which would provide a particularly attractive explanation for the alignment of not only orbital axes but also the rotational axis in the DS~Tucanae system.

A notable limitation of the method of \bmaroon{\citet{Christian22}} is that the longitude of node of planetary orbits cannot be measured with the transit method, which thus limits the measurement of planet-binary mutual inclinations to a minimum value for each \boldnew{system}. Indeed, the only widely applicable exoplanet detection technique that can be used to measure the longitude of node is astrometry. As relatively few exoplanets have been detected \boldnew{with this technique} (both relative, i.e. direct imaging, and absolute \boldnew{astrometry}), and fewer still belong to multiple star systems, it is clearly not possible to conduct a study of planet-binary mutual inclinations based on astrometric data matching the scope of \bmaroon{\citet{Christian22}} at the present time. However, the Gaia mission is expected to detect many thousands of planets with astrometry by the conclusion of its nominal mission \citep{Perryman14}, and this will undoubtedly provide a large sample of planets in binary star systems where it would be possible to measure the planet-binary mutual inclination. We therefore reason that once this sample of astrometrically-detected planets becomes available with future Gaia data releases, it would be highly interesting to investigate the distribution of planet-binary mutual inclinations for a large sample of systems, using similar techniques as applied in this work to the HR~5183-HIP~67291 system. The distribution of planet-binary mutual inclinations afforded by this would allow for indirect constraints on the processes of planet formation that occur in multiple star systems.

\subsection{Stellar multiplicity and the formation of highly eccentric planets} \label{subsec:eccentricity_multiplicity}

\begin{table*}
	\centering
	\caption{Planets with $e\geq 0.8$ in systems without known stellar companions. Masses are minimum masses ($m\sin i$) unless otherwise noted.}
	\label{table:eccentric_planets_nonbinary}
	\begin{tabular}{lcccc}
		\hline
		Planet & Eccentricity & Mass [$M_J$] & Semi-major axis [AU] & Reference \\
		\hline
		HD 219828 c & $0.8102 \pm 0.0051$ & $14.6 \pm 2.3$ & $5.79 \pm 0.41$ & \citet{Ment18} \\
		TOI-3362 b & $0.815^{+0.023}_{-0.032}$ & $5.029^{+0.668}_{-0.646}$ $^{{a}}$ & $0.153^{+0.002}_{-0.003}$ & \citet{Dong21} \\
		HD 22781 b & $0.8191 \pm 0.0023$ & $13.65 \pm 0.97$ & $1.167 \pm 0.039$ & \citet{Diaz12} \\
		HD 43197 b & $0.83^{+0.05}_{-0.01}$ & $0.60^{+0.12}_{-0.04}$ & $0.92^{+0.01}_{-0.02}$ & \citet{Naef10} \\
		Kepler-419 b & $0.833 \pm 0.013$ & $2.5 \pm 0.3$ $^{{a}}$ & $0.370^{+0.007}_{-0.006}$ & \citet{Dawson14} \\
		Kepler-1656 b & $0.836^{+0.013}_{-0.012}$ & $0.153^{+0.013}_{-0.012}$ $^{{a}}$ & $0.197 \pm 0.021$ & \citet{Brady18} \\
		WASP-53 c & $0.8369^{+0.0069}_{-0.0070}$ & $>16.35^{+0.86}_{-0.82}$ & $>3.73^{+0.16}_{-0.14}$ & \citet{Triaud17} \\
		HD 98649 b & $0.852^{+0.033}_{-0.022}$ & $9.7^{+2.3}_{-1.9}$ $^{{a}}$ & $5.97^{+0.24}_{-0.21}$ & \citet{Li21} \\
		HD 76920 b & $0.8782 \pm 0.0025$ & $3.13^{+0.41}_{-0.43}$ & $1.090^{+0.068}_{-0.077}$ & \citet{Bergmann21} \\
		Kepler-1704 b & $0.921^{+0.010}_{-0.015}$ & $4.15 \pm 0.29$ $^{{a}}$ & $2.026^{+0.024}_{-0.031}$ & \citet{Dalba21} \\
		\hline
		\multicolumn{4}{l}{$^{{a}}$ True mass $m$ rather than $m\sin i$.} \\
	\end{tabular}
\end{table*}

\begin{table*}
	\centering
	\caption{Planets with $e\geq 0.8$ that are in known multi-star systems. Masses are minimum masses unless otherwise noted; binary separations are projected values except when otherwise indicated.}
	\label{table:eccentric_planets_binary}
	\begin{tabular}{lcccccc}
		\hline
		Planet & Eccentricity & Mass [$M_J$] & Semi-major axis [AU] & Planet Reference & Binary separation [AU] & Binary Reference \\
		\hline
		HD 7449 b & $0.80^{+0.08}_{-0.06}$ & $1.09^{+0.52}_{-0.19}$ & $2.33^{+0.01}_{-0.02}$ & \citet{Rodigas16} & 21 & \citet{Rodigas16} \\
		HD 28254 b & $0.81^{+0.05}_{-0.02}$ & $2.15^{+0.04}_{-0.05}$ & $2.15^{+0.04}_{-0.05}$ & \citet{Naef10} & 269 & \citet{ElBadry21} \\
		HD 26161 b & $0.820^{+0.061}_{-0.050}$ & $13.5^{+8.5}_{-3.7}$ & $20.4^{+7.9}_{-4.9}$ & \citet{Rosenthal21} & 561 & \citet{ElBadry21} \\
		HD 108341 b & $0.85^{+0.09}_{-0.08}$ & $3.5^{+3.4}_{-1.2}$ & $2.00 \pm 0.04$ & \citet{Moutou15} & 382 & \citet{ElBadry21} \\
		HD 156846 b & $0.84785 \pm 0.00050$ & $10.57 \pm 0.29$ & $1.096 \pm 0.021$ & \citet{Kane11} & 250 & \citet{Tamuz08} \\
		HD 80869 b & $0.862^{+0.028}_{-0.018}$ & $4.86^{+0.65}_{-0.29}$ & $2.878^{+0.045}_{-0.046}$ & \citet{Demangeon21} & 250 & \citet{ElBadry21} \\
		\textbf{HR 5183 b} & $0.87 \pm 0.04$ & $3.31^{+0.18}_{-0.14}$ $^{{a}}$ & $22.3^{+11.0}_{-5.3}$ & This Work & 15400 & This Work \\
		BD+63 1405 b & $0.88 \pm 0.02$ & $3.96 \pm 0.31$ & $2.06 \pm 0.14$ & \citet{Dalal21} & 97 & \citet{ElBadry21} \\
		HD 4113 b & $0.8999^{+0.0020}_{-0.0016}$ & $1.602^{+0.076}_{-0.075}$ & $1.298 \pm 0.030$ & \citet{Cheetham18} & $23.0^{+4.0}_{-2.7}$ $^{{b}}$ & \citet{Cheetham18} \\
		HD 80606 b & $0.93226^{+0.00064}_{-0.00069}$ & $4.116^{+0.097}_{-0.100}$ $^{{a}}$ & $0.4565^{+0.0051}_{-0.0053}$ & \citet{Bonomo17} & 1355 & \citet{ElBadry21} \\
		HD 20782 b & $0.950 \pm 0.001$ & $1.488^{+0.105}_{-0.107}$ & $1.365^{+0.047}_{-0.050}$ & \citet{Udry19} & 9075 & \citet{ElBadry21} \\
		\hline
		\multicolumn{7}{l}{$^{{a}}$ True mass rather than $m\sin i$ $^{{b}}$ Semi-major axis based on an orbital model. As well as a brown dwarf-mass companion, HD 4113 also has a M0-1 stellar} \\
		\multicolumn{7}{l}{companion at a projected separation of 2157 AU \citep{Mugrauer14}.} \\
	\end{tabular}
\end{table*}

While HR~5183~b has one of the most eccentric orbits among known planets, it is hardly alone in this area of the parameter space. At the time of writing, a query to the NASA Exoplanet Archive\footnote{\url{https://exoplanetarchive.ipac.caltech.edu/}, accessed 2021-10-25.} for planets with $e\geq 0.8$ returns 21 results, of which nine were discovered subsequent to 2015 and five were added to this group during 2021 alone. These planets are found at a wide range of orbital periods, from the 18.1-day period of TOI-3362~b \citep{Dong21} to the $\sim$100-year period of HR~5183~b. Most of these planets are of Jovian mass or above, with the least massive exemplar known being Kepler-1656~b with a mass of $0.153^{+0.013}_{-0.012}$ $M_J$ \citep{Brady18}.

Although previous studies have explored the evolution of single members of this high-eccentricity planet sample \citep[e.g.][]{Dawson14, Santos16, Mustill22, Dong21}, few have previously considered this population as a whole. Considering that a substantial number of planets with $e\geq 0.8$ are now known, it seems possible to begin to consider the origins of these planets and their extreme orbits.

One factor that is particularly worth considering in this context is the role of stellar multiplicity. It has been amply theoretically demonstrated that gravitational interactions with a wide companion can dynamically drive a planet towards high orbital eccentricities \citep{Holman97, Mazeh97, Wu03, Takeda05}. This occurs as a result of the Kozai-Lidov effect \citep{Kozai62, Lidov62}, which causes the eccentricity and mutual inclination of the planet to oscillate on long time-scales. \citet{Blunt19} justifiably expressed scepticism that dynamical interactions with HIP~67291 could have played a role in the origin of HR~5183~b's high orbital eccentricity, owing to the wide projected separation of the binary (15400 AU). However, the results of \citet{Mustill22} suggest that this is indeed plausible, in part due to the fact that the periastron distance of the binary is likely to be significantly smaller than the current projected separation (see Section \ref{subsec:binary_results}). To this point, it is important to note that the orbits of wide binaries vary over time due to stellar flybys and interactions with the Galactic tide \citep{Kaib13, CorreaOtto17.perturbations, Pearce21, Mustill22}, such that even binaries with large semi-major axes ($\sim$10$^4$~AU) may undergo periods where the periastron distance reaches as low as $\sim$100~AU, where dynamical interactions with inner planetary systems are greatly amplified \citep{Kaib13, CorreaOtto17.potential}. Thus, the importance of the Kozai-Lidov effect in exciting planetary eccentricities cannot be discounted, even for very widely separated multi-star systems.

However, \citet{Carrera19} have found that planet-planet scattering can also produce planets with such high eccentricities, and have furthermore cast doubt on the role of the Kozai-Lidov effect in the formation of high-eccentricity planets by pointing out that such planets observed in wide binaries could be the result of planet-planet scattering, this itself possibly being initiated by the Kozai-Lidov effect \citep[e.g.][]{Malmberg07, Mustill17}. Indeed, while the results of \citet{Mustill22} indicate that Kozai-Lidov oscillations caused by HIP~67291 can reproduce the high eccentricity of HR~5183~b, it is substantially more probable that the Kozai-Lidov effect can generate the observed eccentricity if planet-planet scattering occurs first than if it acts on a lone planet. While these results may reduce the importance of the Kozai-Lidov effect in the formation of highly eccentric planets in favour of planet-planet scattering, it is important to distinguish this from the role of stellar multiplicity itself; it remains possible that the presence of a stellar companion can cause a planet-planet scattering event that in turn results in the formation of a high-eccentricity planet.

Regardless of whether the Kozai-Lidov effect plays a direct or indirect role, if the presence of a stellar companion significantly contributes to the formation of high-eccentricity exoplanets then these planets would be expected to be more commonly found in multi-stellar systems. \citet{Kaib13} have observed this effect among the general population of exoplanets, finding that the eccentricity distribution of planets in binary systems is skewed towards higher values. However the inverse hypothesis, where high-eccentricity planets are preferentially found in multi-star systems, has not yet been been demonstrated. \citet{Mustill22} observed some evidence for this in systems containing exoplanets with $e\geq 0.8$, but did not go so far as to evaluate the statistical significance of this result. We therefore aim to fully evaluate this hypothesis in this study.

Starting with the 21 known systems containing planets with eccentricities above $e\geq 0.8$, we have searched the astronomical literature for evidence of stellar companions. We have made particular use of the binary catalogue of \citet{ElBadry21}, which is based on Gaia EDR3 \citep{GaiaEDR3} and contains many binaries that have not been previously recognised elsewhere. We present our results in Table \ref{table:eccentric_planets_nonbinary} (for single-star systems) and Table \ref{table:eccentric_planets_binary} (for multi-star systems), with both tables reporting the planetary eccentricities, masses, and semi-major axes, and \bmaroon{the latter} additionally reporting the binary projected separations.

We identify a total of 11 multiple star systems in our sample. All of these systems \bmaroon{are binaries}, although as well as a stellar companion at a projected separation of 2157~AU HD~4113 contains a brown dwarf companion with a semi-major axis of $23.0^{+4.0}_{-2.7}$~AU \citep{Cheetham18, Mugrauer14}, the closer brown dwarf companion being more relevant dynamically to \bmaroon{the interior planet}. The stellar companions in these systems are found at a wide range of separations, from the 21~AU projected separation of HD~7449~B \citep{Rodigas16} to the 15400~AU projected separation of HIP~67291 from HR~5183.

The 11 binary systems in our sample of 21 targets results in a multiplicity rate of $52 \pm 16\%$ (where the reported uncertainty is Poisson noise). To assess whether this multiplicity fraction differs significantly from the overall exoplanet population, we must compare our result with those of a larger sample. For this purpose we make use of the recent results of \citet{Fontanive21}, who conducted a census of multiplicity of exoplanet host stars within 200 pc. The authors measured an overall raw multiplicity rate of $23.2 \pm 1.6\%$, and have also provided counts arranged by certain parameters, one of which is mass; these have been split between <0.1~$M_J$, $0.1-7$~$M_J$, and >7~$M_J$. Of the 21 planets in our $e\geq 0.8$ sample all have masses above >0.1~$M_J$, so to derive a multiplicity rate more comparable to our sample we combine the system counts belonging to the higher-mass bins of \citet{Fontanive21}, resulting in an adopted multiplicity fraction of $25.5 \pm 1.8\%$ for systems including giant planets (>0.1~$M_J$).

While this appears to suggest that our high-eccentricity sample has a higher rate of stellar multiplicity by a factor of $\sim$2, that this is a causal relationship should not yet be considered demonstrated because it cannot be taken for granted that the sample of $e\geq 0.8$ planets is drawn from the same distribution as that of the overall giant planet population. To confirm that the planetary eccentricity does play a causative role, we aim to construct a control sample of planets with similar overall parameters to those in our target group, except in having lower orbital eccentricities. To accomplish this we use a methodology adapted from \bmaroon{\citet{Christian22}}.

To assemble our control sample, we first downloaded the confirmed planet table from the NASA Exoplanet Archive and removed all planets with $e\geq 0.5$. From this table, we matched the three planets that provide the lowest values for the following metric:

\begin{equation}
\left(\frac{a_c-a_p}{a_p}\right)^2 + \left(\frac{m_c-m_p}{m_p}\right)^2 + \left(\frac{M_c-M_p}{M_p}\right)^2 + \log_{10}\left(\frac{\text{distance}_c}{\text{distance}_p}\right)
\:,
\end{equation}

Where $a$ is the planetary semi-major axis in AU, $m$ is the planetary (minimum) mass in $M_J$, $M$ is the stellar mass in $M_\odot$, and distances are in parsecs; terms in $X_p$ reflect the parameters of our 21 target systems and planets, while terms in $X_c$ are as so for the control sample. In this way, we construct a control sample of planets with similar masses, semi-major axes, parent star masses, and distances, but crucially with lower eccentricities than our target planets. Since the observational biases that underlie exoplanet discovery differ significantly between detection methods, we match the discovery techniques used to detect our target planets with those of the control planets; thus, the four planets discovered through the transit method in our sample (TOI-3362~b and Kepler-419~b, -1656~b, -1704~b) were matched only with planets discovered via the transit method, and the remaining target planets were matched with planets discovered through the radial velocity method.

With three control matches for every target planets, we assemble a list of 63 planet matches in the control sample. However, a total of 8 planets are duplicated matches among different target planets, leaving us with a final count of 55 control planets in the same number of systems. We then searched for stellar companions to these systems in the same manner as conducted for the target sample. Our results for the control sample are provided in Appendix \ref{appendix:data}. 13 control systems are identified as stellar multiples, resulting in a multiplicity fraction of $24 \pm 7\%$ for the control sample. This is entirely consistent with our adopted $25.5 \pm 1.8\%$ multiplicity fraction for systems with giant planets. This allows us to securely establish that the high $52 \pm 16\%$ multiplicity rate for our target sample is causally related to the high planetary eccentricities used to select that sample.

We assess the statistical significance of this overabundance of multi-star systems by conducting a Monte Carlo simulation where we assess the probability that a 21-system sample will contain $\geq$11 multi-star systems given an underlying multiplicity fraction of $25.5 \pm 1.8\%$. As a sanity check we also employ a binomial probability test with the same parameters except for a fixed multiplicity fraction of $25.5\%$. Both of these tests result in a probability of $p=0.0075$, suggesting that the rate of stellar multiplicity for our target sample differs from that of the overall giant planet population at a moderate $\sim$2.4$\sigma$ confidence level. We thus conclude that the factor of $\sim$2 overabundance of stellar multiples among high-eccentricity planet hosts is likely to be a real effect, although a larger sample will be required to demonstrate this beyond doubt. 

While this result suggests that the Kozai-Lidov effect plays an important role in the formation of planets with strongly excited orbital eccentricities, it should be remembered that this does not necessarily imply that the Kozai-Lidov effect is the sole mechanism that leads to the formation of highly eccentric planets, as the role of planet-planet scattering cannot be discounted even in binary systems where the Kozai-Lidov effect is likely to occur \citep[hence][]{Carrera19}. However, even if one does not allow a commanding role to the Kozai-Lidov effect, our result indicates that its role in the formation of high-eccentricity planets should not be lightly disregarded. Indeed, it appears likely that the Kozai-Lidov effect plays a significant role in the excitation of planetary eccentricities even in binary systems with very wide separations. In addition to the HR~5183-HIP~67291 system studied in this work with a 15400~AU binary projected separation, it is remarkable that the most eccentric planet currently known, HD~20782~b \citep[$e=0.950 \pm 0.001$,][]{Udry19}, is also found in a very wide binary with a projected separation of 9075~AU. While it is not unlikely that the stellar companions in these systems may approach to significantly smaller separations during periastron, thus producing stronger dynamical perturbations than at their current separations, it nevertheless remains the case that the dynamical influence from a stellar companion should not be discounted in its role in exciting planetary orbits due to a large binary separation alone.

\section{Conclusions} \label{sec:conclusions}

In this work we have performed a new analysis of the HR~5183 system. We have used Hipparcos-Gaia astrometry \citep{Brandt18, Brandt21} to measure the orbital inclination of the highly eccentric planet HR~5183~b discovered by \citet{Blunt19}, and have found that the planet is observed at $i=89.9^{+13.3}_{-13.5}\degree$, consistent with an edge-on orbit. We confirm the previously reported long orbital period and high eccentricity of HR~5183~b, finding $P=102^{+84}_{-34}$~years and $e=0.87 \pm 0.04$. We have furthermore found that HIP~67291 almost certainly forms a physical binary with HR~5183 with a projected separation of 15400~AU, and have used LOFTI \citep{Pearce20} to explore the orbit of this pair. While the uncertainties on the orbital parameters are significant, we find a most probable binary semi-major axis of \boldnew{$\sim$9000~AU} and orbital period of \boldnew{$\sim7\times10^5$~years, with a strong preference for highly inclined orbits ($i\approx80\degree$)}. Combining our constraints on the orbital inclinations and longitudes of node for the planetary and binary orbits, we attempt the mutual inclination between the two orbits in the system \boldnew{for the first time;} we observe a preference for planet-binary misalignment\boldnew{ ($\Delta i>13\degree$ at 3$\sigma$ confidence), which remains significant even after acknowledging that the mutual inclination prior is biased towards misaligned orbits. However, our measurement of the misalignment of orbits in the system is insufficiently precise to make use of the hypothesis of \citet{Mustill22} that the planet-binary mutual inclination reflects the formation history or HR~5183~b.} Future observations will allow for the mutual inclination in this system to be more precisely constrained, possibly allowing for direct constraints on the formation of HR~5183~b.

Of the 21 systems containing planets with eccentricities above $e\geq 0.8$ currently known, 11 are found in multi-stellar systems. This rate of multiplicity exceeds that of the overall planet-host population by a factor of $\sim$2, a result that we demonstrate to be moderately significant ($p=0.0075$). This provides observational support for the hypothesis that dynamical interactions with exterior stellar companions through the Kozai-Lidov effect plays a major role in the formation of highly eccentric exoplanets, although planet-planet scattering is likely to be an important factor as well. A larger sample of systems with highly eccentric exoplanets will be required before this overabundance of stellar multiples can be demonstrated beyond doubt.

A growing number of long-period exoplanets have been detected with Hipparcos-Gaia astrometry to date \citep[e.g.][]{Feng19, Xuan20, Li21}, providing new insights into the population of giant planets discovered by the radial velocity method. This group of detections will however undoubtedly be dwarfed by the number of planets that will be found in the Gaia DR4 astrometric solution \citep{Perryman14}. Of the science discoveries that could be accomplished with this sample, we propose that the planet population revealed by Gaia will allow for the study of mutual inclinations for planets in wide binary systems to an extent that has not previously been possible, through the use of techniques similar to those utilised in this study. The distribution of planet-binary mutual inclinations afforded by this could be used to constrain the planet formation processes that occur in multiple star systems.

\section*{Acknowledgements}

\bmaroon{We thank the referee for many helpful suggestions that have improved this manuscript. We thank Alex Mustill \boldnew{and Timothy D. Brandt} for helpful comments.} The authors would like to thank Sarah Blunt, Jerry W. Xuan, and Judah van Zandt for their courtesy and co-operation during the inception of this project. This research has made use of the SIMBAD database and VizieR catalogue access tool, operated at CDS, Strasbourg, France. This research has made use of NASA's Astrophysics Data System. \boldagain{This research has made use of the NASA Exoplanet Archive, which is operated by the California Institute of Technology, under contract with the National Aeronautics and Space Administration under the Exoplanet Exploration Program.} This work has made use of data from the European Space Agency (ESA) mission {\it Gaia} (\url{https://www.cosmos.esa.int/gaia}), processed by the {\it Gaia} Data Processing and Analysis Consortium (DPAC, \url{https://www.cosmos.esa.int/web/gaia/dpac/consortium}). Funding for the DPAC has been provided by national institutions, in particular the institutions participating in the {\it Gaia} Multilateral Agreement.

%%%%%%%%%%%%%%%%%%%%%%%%%%%%%%%%%%%%%%%%%%%%%%%%%%
\section*{Data Availability}

The radial velocity data used in this article are found in \citet{Blunt19} (\url{https://vizier.cds.unistra.fr/viz-bin/VizieR?-source=J/AJ/158/181}). The astrometric data used here are found in \citet{GaiaEDR3} (\url{https://vizier.cds.unistra.fr/viz-bin/VizieR?-source=I/350}) and \citet{Brandt21} (\url{https://doi.org/10.3847/1538-4365/abf93c}).
 
%The inclusion of a Data Availability Statement is a requirement for articles published in MNRAS. Data Availability Statements provide a standardised format for readers to understand the availability of data underlying the research results described in the article. The statement may refer to original data generated in the course of the study or to third-party data analysed in the article. The statement should describe and provide means of access, where possible, by linking to the data or providing the required accession numbers for the relevant databases or DOIs.



\bibliographystyle{mnras}
\bibliography{bib} % if your bibtex file is called example.bib

%%%%%%%%%%%%%%%%%%%%%%%%%%%%%%%%%%%%%%%%%%%%%%%%%%

%%%%%%%%%%%%%%%%% APPENDICES %%%%%%%%%%%%%%%%%%%%%

\appendix

\section{Supporting Figures} \label{appendix:figures}

\begin{figure*}
	\includegraphics[width=0.95\textwidth]{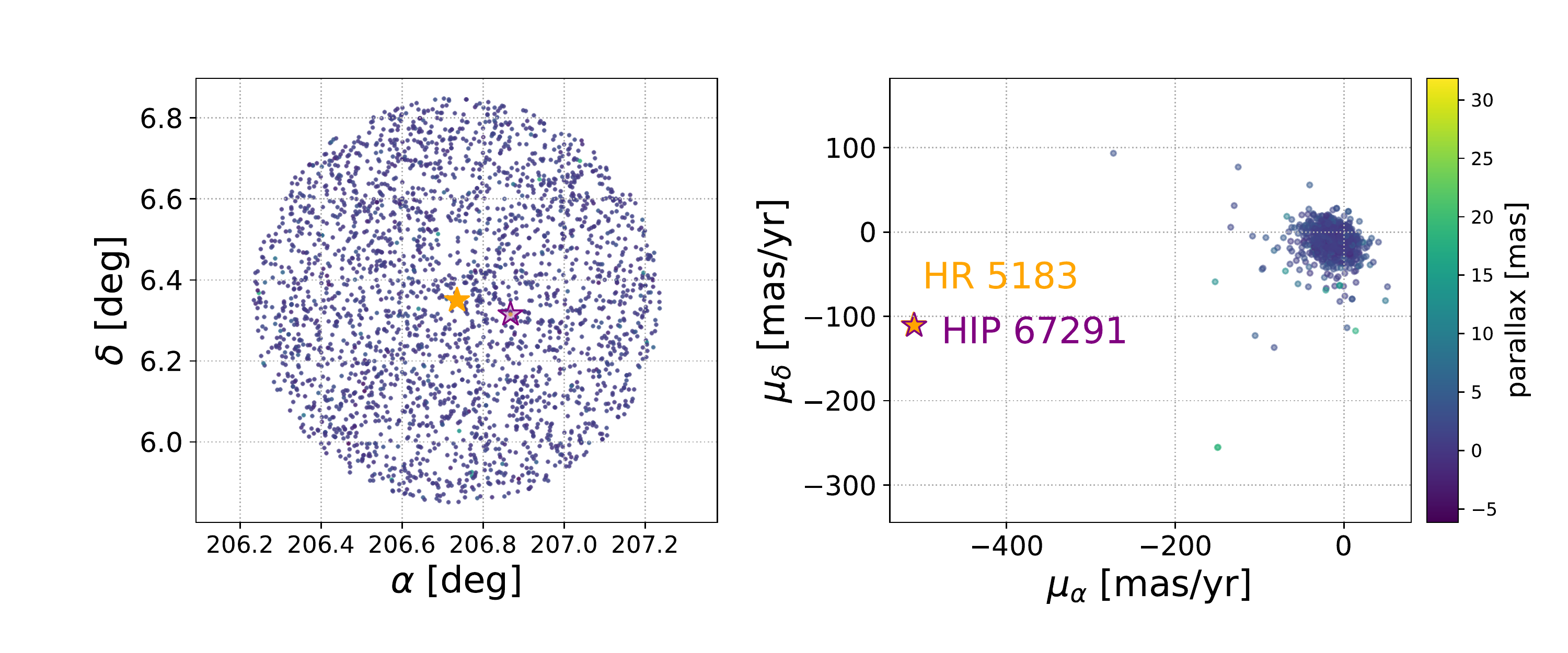}
	\caption{Gaia EDR3 sources within a 0.5 degree (30 arcminute) radius of HR~5183, colour-coded by parallax, in position (left) and proper motion (right). HR~5183 is marked by a filled yellow star while HIP~67291 is marked by an unfilled purple star. These figures demonstrate the low probability that HIP~67291 is an unrelated star observed in a chance alignment with HR~5183, as discussed in Section \ref{subsec:binary_method}. Note that the right panel is very similar to figure 1 in \citet{Mustill22}, who used a smaller 20 arcminute radius to derive comparable results.}
	\label{figure:appendix_fig}
\end{figure*}

In Figure \ref{figure:appendix_fig} we show the results of a query to Gaia EDR3 for sources found within 0.5 degrees of HR~5183, as was used in Section \ref{subsec:binary_method} to determine the probability that HIP~67291 is an unrelated star observed in a chance alignment. The close agreement in parallaxes and proper motions between these two stars strongly argues against this hypothesis. Note that the right panel of Figure \ref{figure:appendix_fig} is very similar to figure 1 in \citet{Mustill22}.

\section{Supporting Data} \label{appendix:data}

\begin{table*}
	\centering
	\caption{Our control sample used in Section \ref{subsec:eccentricity_multiplicity}. \bmaroon{Duplicated} matches are indicated by parentheses.}
	\label{table:eccentric_planets_control}
	\begin{tabular}{lccccccc}
		\hline
		Target planet & Matched planet 1 & Multiple? & Matched planet 2 & Multiple? & Matched planet 3 & Multiple? & References \\
		\hline
		HD 219828 c & WASP-8 c & Yes & HD 115954 b & No & HD 38529 c & Yes & \citet{Queloz10, Raghavan06} \\
		TOI-3362 b & TOI-558 b & No & TOI-172 b & No & WASP-162 b & No &  \\
		HD 22781 b & HD 141937 b & No & HD 28185 b & No & HD 114762 b & Yes & \citet{Bowler09} \\
		HD 43197 b & BD+55 362 b & No & HD 43564 c & No & HD 164509 b & Yes & \citet{Ngo17} \\
		Kepler-419 b & Kepler-30 c & No & Kepler-117 c & No & Kepler-1657 b & No &  \\
		Kepler-1656 b & K2-280 b & No & TOI-216 c & No & HD 221416 b & No &  \\
		WASP-53 c & HATS-59 c & No & HD 214823 b & Yes & Kepler-129 d & No & \citet{Mugrauer19} \\
		HD 98649 b & (WASP-8 c) &  & (HD 115954 b) &  & HD 30177 b & No &  \\
		HD 76920 b & WASP-41 c & No & HD 73526 c & No & 42 Dra b & Yes & \citet{Mugrauer19} \\
		Kepler-1704 b & Kepler-1514 b & No & (Kepler-1657 b) &  & HATS-61 b & Yes & \citet{ElBadry21} \\
		HD 7449 b & HD 141399 d & No & $\tau^1$ Gru b & No & HD 187085 b & No &  \\
		HD 28254 b & (HD 141399 d) &  & HD 90156 b & No & HD 108874 c & No &  \\
		HD 26161 b & HD 68988 c & No & HD 92987 b $^a$ & No & (HD 30177 c) &  &  \\
		HD 108341 b & HD 73267 b & No & HD 143361 b & No & HD 68402 b & No &  \\
		HD 156846 b & $\gamma^1$ Leo b & Yes & HD 33564 b & No & 4 UMa b & No & \citet{Han10} \\
		HD 80869 b & HD 11506 b & No & HIP 8541 b & Yes & HD 132406 b & No & \citet{ElBadry21} \\
		HR 5183 b & HD 92788 c & No & HD 50499 c & No & HD 150706 b & No &  \\
		BD+63 1405 b & (HD 73267 b) &  & HD 143361 b & No & HD 128311 c & No &  \\
		HD 4113 b & HD 188015 b & Yes & HD 133131 Ab & Yes & HD 142415 b & No & \citet{Raghavan06, Teske16} \\
		HD 80606 b & HD 35759 b & No & HD 12484 b & Yes & HD 32518 b & No & \citet{Shaya11} \\
		HD 20782 b & (HD 133131 Ab) &  & (HD 188015 b) &  & HD 65216 b & Yes & \citet{Mugrauer07} \\
		\hline
		\multicolumn{8}{l}{$^a$ At the time of our analysis HD 92987 b was listed in the NASA Exoplanet Archive confirmed planet table. However, \citet{Venner21} have demonstrated that} \\
		\multicolumn{8}{l}{this companion is actually a star observed at a near-polar orbital inclination, and the companion has subsequently been removed from the table.} \\
	\end{tabular}
\end{table*}

In Table \ref{table:eccentric_planets_control} we list our control sample of low-eccentricity planets used in Section \ref{subsec:eccentricity_multiplicity}. The planets from our target sample (Tables \ref{table:eccentric_planets_nonbinary}, \ref{table:eccentric_planets_binary}) are listed in the first column, followed by the three planets that provide the best match to their parameters and our assessment of the stellar multiplicity in their systems. Duplicated system matches are indicated by parentheses and are not counted in our statistics. See Section \ref{subsec:eccentricity_multiplicity} for further details.

%%%%%%%%%%%%%%%%%%%%%%%%%%%%%%%%%%%%%%%%%%%%%%%%%%

% Don't change these lines
\bsp	% typesetting comment
\label{lastpage}
\end{document}